\documentclass[%
reprint,
superscriptaddress,
amsmath,
amssymb,
aps,
prb,
floatfix,
longbibliography
]{revtex4-1}
\usepackage{amssymb}
\usepackage{graphicx}
\usepackage{appendix}
\usepackage{color}
\usepackage{subfigure}
\usepackage[colorlinks=true,linkcolor=blue,filecolor=blue, urlcolor=blue,citecolor=blue]{hyperref}

\begin{document}

\preprint{APS/123-QED}

\graphicspath{{mainfigures/}}
\title{Spontaneous charge-ordered state in Bernal-stacked bilayer graphene}
\author{Xiu-Cai Jiang}
\affiliation{Shanghai Key Laboratory of Special Artificial Microstructure Materials and Technology, School of Physics Science and engineering, Tongji University, Shanghai 200092, People’s Republic of China}
\author{Ze-Yi Song}
\affiliation{Shanghai Key Laboratory of Special Artificial Microstructure Materials and Technology, School of Physics Science and engineering, Tongji University, Shanghai 200092, People’s Republic of China}
\author{Ze Ruan}
\affiliation{Shanghai Key Laboratory of Special Artificial Microstructure Materials and Technology, School of Physics Science and engineering, Tongji University, Shanghai 200092, People’s Republic of China}
\author{Yu-Zhong Zhang}
\email[Corresponding author.]{Email: yzzhang@tongji.edu.cn}
\affiliation{Shanghai Key Laboratory of Special Artificial Microstructure Materials and Technology, School of Physics Science and engineering, Tongji University, Shanghai 200092, People’s Republic of China}

\date{\today}

\begin{abstract}
We propose that a weakly spontaneous charge-ordered insulating state probably exists in Bernal-stacked bilayer graphene which can account for experimentally observed non-monotonic behavior of resistance as a function of the gated field, namely, the gap closes and reopens at a critical gated field. The underlying physics is demonstrated by a simple model on a corresponding lattice that contains the nearest intralayer and interlayer hoppings, electric field, and staggered potential between different sublattices. Combining density functional theory calculations with model analyses, we argue that the interlayer van der Waals interactions cooperating with ripples may be responsible for the staggered potential which induces a charge-ordered insulating state in the absence of the electric field.
\end{abstract}

\maketitle
\section{Introduction}
Bilayer graphene has been extensively studied for more than one decade. While fascinating progresses, such as discovering of the integer quantum Hall effect~\cite{novoselov2006unconventional,kou2014electron}, superconductivity~\cite{cao2018unconventional,lu2019superconductors}, higher-order topological insulators~\cite{park2019higher}, tunable excitons~\cite{park2010tunable,ju2017tunable}, and topological valley transport~\cite{sui2015gate}, induced by external fields~\cite{rozhkov2016electronic,oostinga2008gate,feldman2009broken,taychatanapat2010electronic,maher2013evidence,li2019metallic}, doping~\cite{ohta2006controlling,park2012single}, and twist~\cite{mele2010commensuration,shallcross2010electronic,cao2018correlated}, have been reported, the ground state of Bernal-stacked bilayer graphene (BBG) remains controversial. Originally, BBG was thought to be a semimetal with massive Dirac cones at the Fermi level. And a gated field applied perpendicular to the plane breaks the symmetry between the top and bottom layers of Bernal-stacked bilayer graphene (BBG), rendering it an insulator, which has been confirmed by transport~\cite{oostinga2008gate,taychatanapat2010electronic} and photoemission experiments~\cite{zhang2009direct,mak2009observation}. Then, the gap should increase monotonously as a function of the applied gated electric field~\cite{mccann2007low,PhysRevB.74.161403,PhysRevLett.99.216802,mccann2013electronic}.
However, this is challenged by intriguing experimental observations on ultraclean suspended BBG that resistance varies non-monotonously with the gated electric field at low temperatures~\cite{li2019metallic,weitz2010broken,bao2012evidence,velasco2012transport}, suggesting the presence of an intrinsic gap that closes and then reopens when an electric field is applied.

So far, much effort has been made to understand the discrepancy. Starting with the intrinsic gap at zero field, various possible candidate states which stem from different origins have been proposed. Using methods like quantum Monte Carlo, functional renormalization group, etc., a layered antiferromagnetic (LAF) state has been suggested as a candidate state in BBG, which is favored by on-site Coulomb interactions~\cite{lang2012antiferromagnetism,yuan2013possible,jung2011lattice,wang2013layer,scherer2012instabilities}. By calculating the properties of Landau level $n$=0 with spin and valley degree of freedom, a canted antiferromagnetic (CAF) state is suggested to be stabilized by isospin anisotropy of electron-electron and electron-phonon interactions~\cite{kharitonov2012canted}.  Besides, the quantum spin Hall (QSH) state~\cite{scherer2012instabilities,lemonik2012competing} and quantum anomalous Hall (QAH) state~\cite{nandkishore2010quantum} are also considered as two potential candidate states with gap opening at zero field, which are favored by spin-orbit coupling and zero-point fluctuations, respectively. Recently, taking short-range interactions into account, a candidate state with the coexistence of nematic and antiferromagnetic states has also been proposed~\cite{ray2021gross}.

However, despite numerous investigations, a definitive explanation for the phenomenon that resistance varies non-monotonously with an electric field remains elusive, which is probably due to the following two reasons. On one hand, most of the previous studies focus on the ground state at zero field, where there are many competing candidate states with very close energies~\cite{jung2011lattice}. Consequently, the ground state strongly relies on delicate details of the microscopic model~\cite{scherer2012instabilities,xu2016gate,ray2021gross,PhysRevB.86.075467,PhysRevB.86.115447,PhysRevB.103.205135} that a specific perturbation may favor a particular candidate state as introduced above. Although these studies suggest the presence of a magnetic ground state at zero field~\cite{zhang2011spontaneous}, there is no direct experimental evidence, such as a neutron diffraction experiment, for the existence of spontaneous magnetization in BBG. On the other hand, the models employed to investigate the gap include only several tight-binding parameters to describe the low-energy dispersions in the vicinity of the Dirac point~\cite{yuan2013possible,jung2011lattice,kharitonov2012canted}. Consequently, these models fail to capture the realistic behavior of the gap, exhibiting inconsistencies with experimental observations~\cite{yuan2013possible,jung2011lattice,kharitonov2012canted}.

Therefore, it is necessary to investigate the behavior of the gap under an electric field based on a reasonable model to determine the ground state of BBG. Noticeably, some key ingredients such as interlayer van der Waals (VdW) interactions and ripples are often ignored by previous analyses. The interlayer VdW interactions and the ripples which naturally occur in graphene sheets~\cite{stolyarova2007high,meyer2007structure,fasolino2007intrinsic} are crucial to the properties of BBG since interlayer VdW interactions play a dominant role in anchoring the layers at a fixed distance~\cite{zhang2010band} and ripples can drive graphene (including BBG consisting of two layers) into an insulator~\cite{gui2012electronic,lin2015feature}. Furthermore, it has been suggested that a charge-ordered state, which may be favored by the two aforementioned ingredients in BBG, is possible in suspended graphene samples~\cite{jung2009theory}. Thus, taking into account the effect of interlayer VdW interactions and ripples to provide a comprehensive explanation for the field-induced non-monotonic behavior of the gap is an interesting work.

In this paper, we present a novel explanation for the phenomenon observed in BBG that the resistance varies non-monotonically with an applied electric field, corresponding to the gap closes and then reopens with the electric field. We start by demonstrating the underlying physics of the phenomenon using a simple model with staggered potential between inequivalent sublattices on a Bernal-stacked bilayer honeycomb lattice. We reveal that the intrinsic gap at zero field is attributed to the presence of a particular intralayer charge-ordered state which is characterized by an inverted order of the four low-energy bands, where two touched bands shift below the Fermi level while the other two untouched bands move above it [see Fig.~\ref{schematic-diagram}(b)], in contrast to the disordered case, where touched ones meet at the Fermi level forming a massive Dirac cone [see Fig.~\ref{schematic-diagram}(a)]. As an electric field is applied to this charge-ordered state, the upper band of the two touched bands and the lower band of the two untouched bands move towards and then cross each other, resulting in the non-monotonic behavior of the gap at a small electric field. To validate the proposal that this charge-ordered state exists in BBG, we combine density functional theory (DFT) calculations with model analyses to include the effect of the interlayer van der Waals (VdW) interactions and ripples. We find that interlayer VdW interactions, along with ripples effectively generate a staggered potential between inequivalent sublattices in BBG. Most importantly, the experimental phenomenon is successfully reproduced when the strength of interlayer VdW interactions is on the order of $10$ meV. Our results offer a fresh perspective on the field-induced intriguing phenomenon in BBG.

Our paper is structured as follows. In Section~\ref{model-and-method}, we provide a comprehensive description of the model and methods employed in our study. Section~\ref{BLG:results} presents our primary findings. Specifically, we calculate the gaps and intralayer charge disproportionations as functions of the staggered potential. We also examine how the eigenvalues of the Dirac point and low-energy bands vary when an electric field is applied. Additionally, we investigate the evolution of the gap with respect to the electric field when including the effect of interlayer VdW interactions and ripples. Section~\ref{BLG:discussion} presents a detailed discussion, and Section~\ref{BLG:conclusion} concludes our paper with a concise summary.

\section{model and method}\label{model-and-method}
\begin{figure}[htbp]
\includegraphics[width=0.46\textwidth,height=0.24\textwidth]{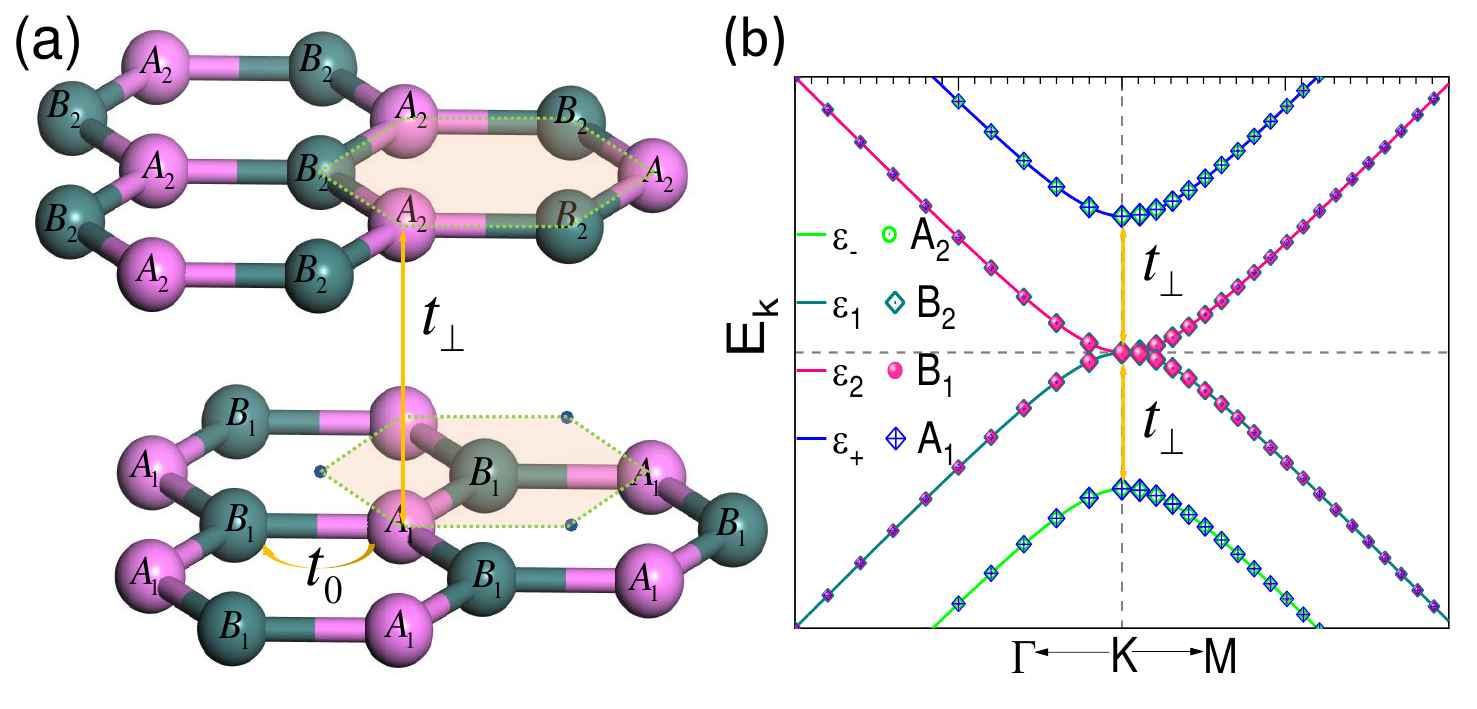}
\caption{(a) The structure of BBG, where the nearest intralayer and interlayer hoppings, namely, $t_{0}$ and $t_{\perp}$ are presented. (b) The low-energy bands of the disordered case, where $H_{\Delta}=0$ and $H_{E}=0$.}
\label{Structure-bands}
\end{figure}

The simple model that we employ to demonstrate the underlying physics for the intriguing phenomenon is given by
\begin{eqnarray}
H=H_{\tau}+H_{K}+H_{\Delta}+H_{E},\label{Total-Hamiltonian}
\end{eqnarray}
where
\begin{eqnarray}
\begin{aligned}
H_{K}&=
-t_{0}\sum\limits_{\langle{ij}\rangle\sigma}\Big[C_{i A_{1}\sigma}^{\dag}C_{jB_{1}\sigma}
+C_{iA_{2}\sigma}^{\dag}C_{jB_{2}\sigma}\Big]+h.c., \\
H_{\tau}&=
-t_{\perp}\sum\limits_{i\sigma}C_{iA_{1}\sigma}^{\dag}C_{iA_{2}\sigma}
-t_{\perp}\sum\limits_{i\sigma}C_{iA_{2}\sigma}^{\dag}C_{iA_{1}\sigma}, \\
H_{\Delta}&=
\frac{1}{2}\Delta\sum\limits_{i\sigma}\sum\limits_{m}{\hat{n}}_{iA_{m}\sigma}
-\frac{1}{2}\Delta\sum\limits_{i\sigma}\sum\limits_{m}{\hat{n}}_{iB_{m}\sigma},\\
H_{E}&=
-\frac{1}{2}Eed\sum\limits_{i\sigma}\sum\limits_{m}(-1)^m\Big[\hat{n}_{iA_{m}\sigma}+\hat{n}_{iB_{m}\sigma}\Big].
\end{aligned}
\end{eqnarray}
Here, $H_{K}$, $H_{\tau}$, $H_{\Delta}$, and $H_{E}$ denote the nearest intralayer hopping terms, the nearest interlayer hopping terms, staggered potential energy, and the external electric field terms, respectively. $C_{iS_m\sigma}$ annihilates an electron with spin $\sigma$ on sublattice $S$ (including A and B) of layer $m$ in $i$th unit cell. $\hat{n}_{iS_m\sigma}$ is the particle-number operator. $\langle{ij}\rangle$ means summation over intralayer nearest-neighbor sites. $t_0$ and $t_{\perp}$ are hopping integrals as depicted in Fig.~\ref{Structure-bands}(a). $\Delta$ is the staggered potential with opposite signs in inequivalent sublattices, arising from distinct atomic environments of $A_{1(2)}$ and $B_{1(2)}$ sublattices, where $A_2$ is on top of $A_1$, whereas $B_2$ ($B_1$) is above (below) the center of the hexagon in the bottom (top) layer. A downward electric field $E$ is studied. $d=3.4$ {\AA} is the interlayer distance, and $e$ is the elementary charge. By applying Fourier transform, the Hamiltonian can be expressed in momentum space. Diagonalizing this Hamiltonian yields the eigenvalues of a given $\boldsymbol{k}$ point. Notably, the eigenvalues of the Dirac point are
\begin{eqnarray}
\begin{aligned}
\varepsilon_{1}&=-\frac{\Delta+eEd}{2},
\quad
\quad
\varepsilon_{-}=\frac{\Delta}{2}-\sqrt{t_{\perp}^{2}+\frac{(eEd)^{2}}{4}},\\
\varepsilon_{2}&=\frac{eEd-\Delta}{2},
\quad
\quad
\varepsilon_{+}=\frac{\Delta}{2}+\sqrt{t_{\perp}^{2}+\frac{(eEd)^{2}}{4}}.
\end{aligned}
\label{dirac-point-E}
\end{eqnarray}
It is easy to find that, $\varepsilon_{1}$ and $\varepsilon_{2}$ are contributed by $p_z$ orbitals of $B_2$ and $B_1$ atoms, respectively, whereas $\varepsilon_{\pm}$ are contributed by a linear combination of $p_z$ orbitals of $A_1$ and $A_2$ atoms.

To accurately reproduce the intriguing phenomenon observed in BBG, it is imperative to include the effect of previously ignored interlayer VdW interactions and ripples. While ripples can be readily introduced through DFT calculations, the inclusion of interlayer VdW interactions is challenging due to the existence of various corrections such as vdW-DF~\cite{dion2004van},
TS-vdW~\cite{tkatchenko2009accurate},
vdW-DF-C~\cite{cooper2010van},
DFT-D~\cite{grimme2010consistent}, etc. Thus, in order to gain insight into the effect of VdW interactions and eliminate the uncertainty from different choices of corrections, we construct a Hamiltonian as
\begin{eqnarray}
H=H_{D}+H_{eff}+H_{E},\label{DFT+vdw+E}
\end{eqnarray}
where
\begin{eqnarray}
H_{D}&=
-\sum\limits_{is\sigma}{\sum\limits_{js^{\prime}}}t_{isjs^{\prime}}C_{is\sigma}^{\dag}C_{js^{\prime}\sigma}
\end{eqnarray}
is the tight-binding Hamiltonian describes the low-energy dispersion of DFT band structures, where the hopping parameters $t_{isjs^{\prime}}$ can be derived through the transformation from Bloch space to maximally localized Wannier functions basis by using WANNIER90 code~\cite{marzari2012maximally,mostofi2008wannier90}.
Four bands close to the Fermi energy which are mainly contributed by $p_z$ orbitals of four carbon sublattices are taken into account. Sufficient numbers of long-range hoppings $t_{isjs^{\prime}}$s are included in order to precisely describe the dispersion of low-energy bands obtained from DFT calculations as shown in Appendix \ref{DFT-fit}. The DFT band structures of BBG are calculated by the full-potential linearized
augmented plane-wave method\cite{singh2006planewaves} within the density functional theory as implemented in the WIEN2K package\cite{blaha2001wien2k}, where The exchange-correlation interactions are treated by the local-density approximation\cite{perdew1981self}. A shifted $60\times60\times1$ special k-point mesh with a modified tetrahedron integration scheme\cite{blochl1994improved} for the sampling of the Brillouin zone is employed. The valence and core states
are separated by an energy of 6.0 Ry and the plane-wave cutoff parameter $R_{mt}\times k_{max}$ is set to be 7.00, where $R_{mt}=1.33$ a.u. is used. The valence wave functions inside the atomic spheres are expanded up to $l_{max}=10$ while the charge density is Fourier expanded up to $G_{max}=12$. We choose both a charge convergence of $10^{-7}$e and an energy convergence of $10^{-7}$ Ry as the convergence criteria.  A sufficiently large vacuum distance
of 17.1 {\AA} is used to eliminate the interactions between periodic images of the layers in the direction perpendicular to the atomic plane. $H_{eff}$ is the effective Hamiltonian capturing the effect of interlayer VdW interactions, and $H_{E}$ is the aforementioned electric field terms.
To obtain the effective Hamiltonian $H_{eff}$, we use a VdW potential of interatomic Lennard–Jones type
\begin{equation}
w_{ij}^{ss^{\prime}}=
V_0\bigg[\Big(\frac{R_{ss^{\prime}}}{|\boldsymbol{r_{is}}-\boldsymbol{r_{js^{\prime}}}|}\Big)^{12}
-2\Big(\frac{R_{ss^{\prime}}}{|\boldsymbol{r_{is}}-\boldsymbol{r_{js^{\prime}}}|}\Big)^{6}\bigg],\label{eqn:atom-VdW}
\end{equation}
where $i(j)$ and $s(s^{\prime})$ correspond to cell and sublattice indices, respectively. $V_0$ is the strength of interlayer VdW interactions, determining the depth of the potential well. $R_{ss^{\prime}}$ represents the bottom of the sublattice-dependent interlayer VdW potential well, including $R_{AA}$, $R_{BB}$, $R_{AB}$, and $R_{BA}$. Since the interlayer VdW interactions play a dominant role in anchoring the layers at a fixed distance~\cite{zhang2010band}, $R_{ss^{\prime}}$ can be approximated by force equilibrium condition of $s$ sublattice in $i$th unit cell
\begin{equation}
12V_0{\sum\limits_{j}}^{\prime}
\Big(\frac{R_{ss^{\prime}}^{6}}{|\boldsymbol{r_{is}}-\boldsymbol{r_{js^{\prime}}}|^{7}}
-\frac{R_{ss^{\prime}}^{12}}{|\boldsymbol{r_{is}}-\boldsymbol{r_{js^{\prime}}}|^{13}}\Big)\hat{\boldsymbol{e}}_{ij}^{ss^{\prime}}=0,\label{eqn:force}
\end{equation}
where $\hat{\boldsymbol{e}}_{ij}^{ss^{\prime}}=(\boldsymbol{r_{is}}-\boldsymbol{r_{js^{\prime}}})/|\boldsymbol{r_{is}}-\boldsymbol{r_{js^{\prime}}}|$ is the unit vector, and $\sum^{\prime}$ represents the summation over the layer without $s$ sublattice. Thus, for a given $V_0$, the potential energy of $s$ sublattices in $i$th unit cell reads
\begin{equation}\label{eqn:VdW}
U_{is}=\frac{1}{2}V_0{\sum\limits_{j}}^{\prime}\sum\limits_{s^{\prime}}w_{ij}^{ss^{\prime}}.
\end{equation}
Interestingly, there is a total potential energy difference between A-type and B-type sublattices as
\begin{equation}\label{deltaU}
\delta{U}_{AB}=\sum\limits_{i}(U_{iA_{1}}+U_{iA_{2}}-U_{iB_{1}}-U_{iB_{2}}),
\end{equation}
namely, the potential energy well of $A_{1(2)}$ is higher than that of $B_{1(2)}$. As a result, electrons redistribute in response to eliminate this difference compared with the case without the interlayer VdW interactions. Thus, interlayer VdW interactions effectively generate a staggered potential between inequivalent sublattices in the layer, namely, $H_{eff}$ satisfies
\begin{equation}
\label{eqn:energydiff}
\begin{aligned}
H_{eff}&=
\Delta\sum\limits_{i\sigma}\sum\limits_{m}\big[{\hat{n}}_{iA_{m}\sigma}
-{\hat{n}}_{iB_{m}\sigma}\big]/2,\\
\delta{U}_{AB}&=-\langle{H_{eff}}\rangle.
\end{aligned}
\end{equation}
Thus, the staggered potential is determined as long as the strength of interlayer VdW interactions $V_0$ is given.

\section{results}\label{BLG:results}
\begin{figure}[htbp]
\includegraphics[width=0.46\textwidth,height=0.38\textwidth]{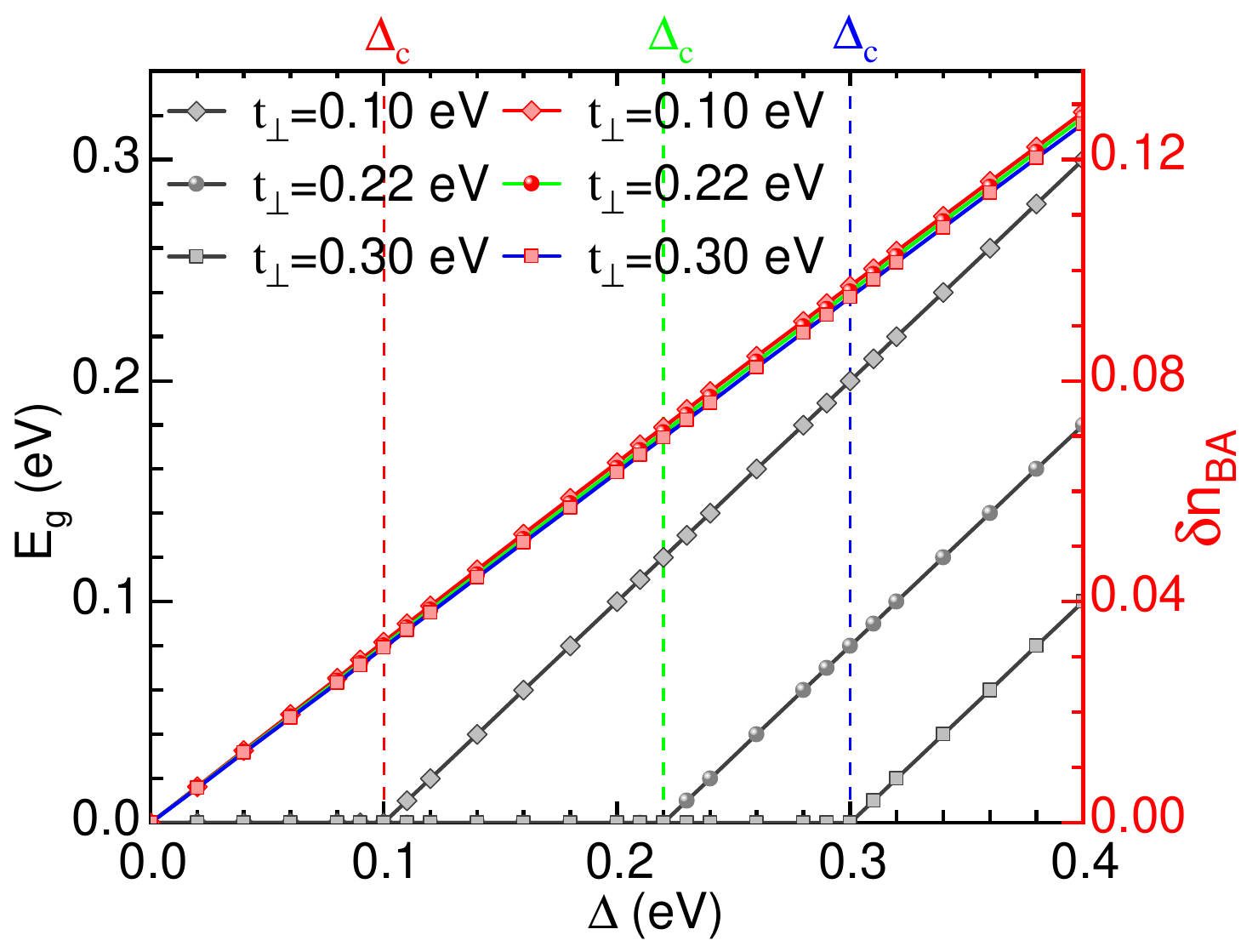}
\caption{The gaps $E_{g}$ and intralayer charge disproportionation $\delta{n}_{BA}$ as functions of staggered potential $\Delta$ for three nearest inter-layer coupling with $t_{\perp}=0.1$ eV, $t_{\perp}=0.22$ eV, and $t_{\perp}=0.3$ eV. $t_0=2.7$ eV is used here. Gray symbols and left axis describe the band gap while red symbols and right axis depict the charge disproportionation $\delta{n}_{BA}$.}
\label{gap-n-delta}
\end{figure}
Here, we demonstrate firstly the underlying physics for the phenomenon that the gap closes and then reopens with the electric field using the simple model~\eqref{Total-Hamiltonian} as introduced above.

\begin{figure}[htbp]
\includegraphics[width=0.48\textwidth,height=0.57\textwidth]{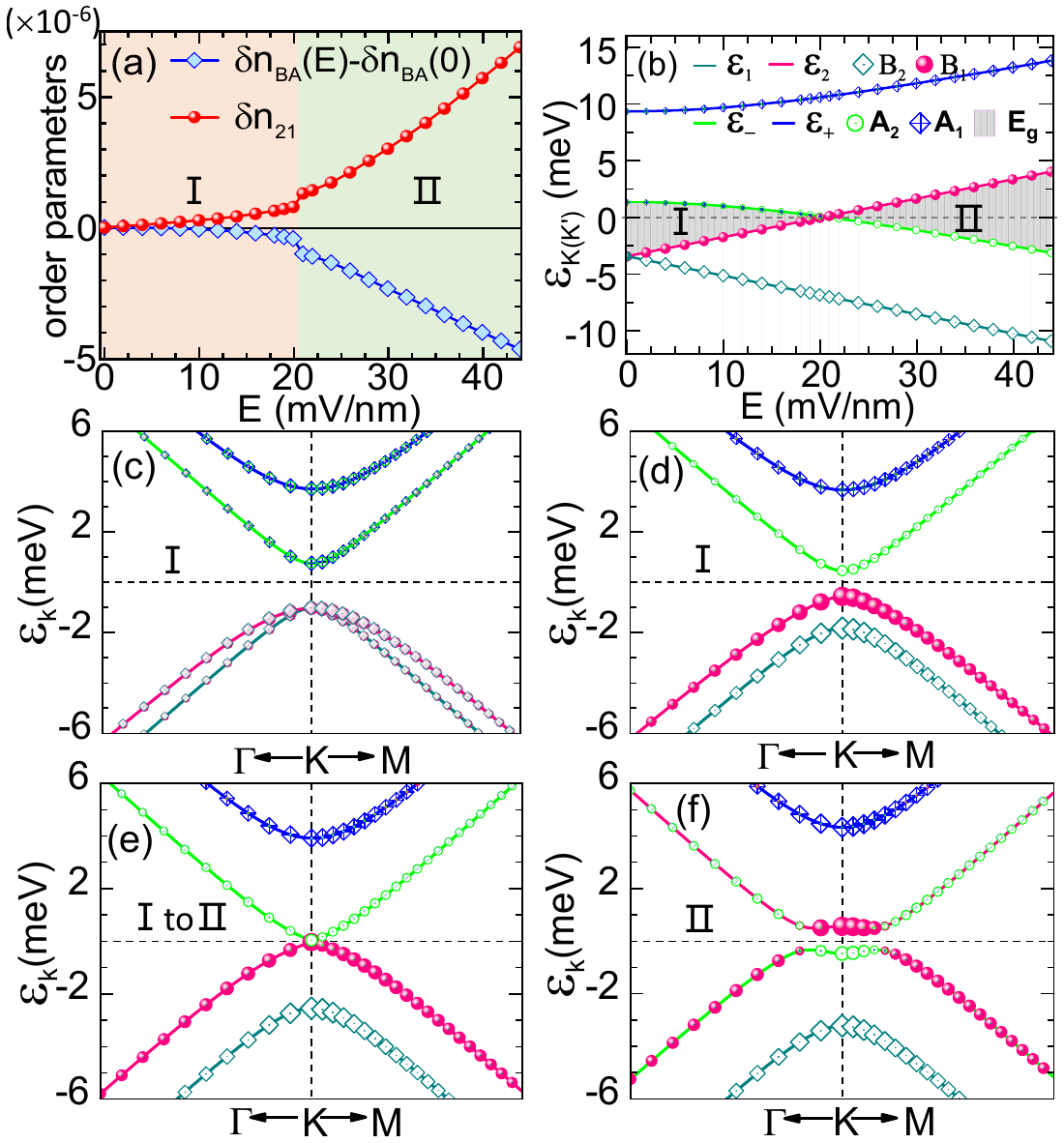}
\caption{(a) $\delta{n}_{21}$ and $\delta{n}_{BA}(E)-\delta{n}_{BA}(0)$ as functions of the electric field, where $\delta{n}_{21}$, $\delta{n}_{BA}(E)$, and $\delta{n}_{BA}(0)$ are the order parameters of interlayer CD, field-dependent intralayer CD, and zero-field intralayer CD, respectively. (b) The field-dependent eigenvalues at the Dirac point.  (c)-(f) present the evolution of low-energy bands with respect to the electric field, including $E=0$ mV/nm (a), $10$ mV/nm (b), $20$ mV/nm (c), and $30$ mV/nm (d). To show all of the low-energy bands within the same energy scale, we use $t_{\perp}=4$ meV, $t_0=2.7$ eV, $\Delta=8.8$ meV here.}
\label{gap-band-field}
\end{figure}

Starting with the insulating state at zero field ($E=0$), we find that the intrinsic gap at zero field is due to the presence of a particular intralayer charge-ordered state where sublattice B is charge-rich while sublattice A is charge-poor. To illustrate this, we calculate the gaps $E_g$ and intralayer charge disproportionation (CD) $\delta{n}_{BA}=\frac{1}{2}(n_{B_1}$+$n_{B_2}$-$n_{A_1}$-$n_{A_2})$ as functions of the staggered potential $\Delta$ for different interlayer couplings $t_{\perp}$ at zero field, as shown in Fig.~\ref{gap-n-delta}. As can be seen, the model always predicts a charge-ordered insulating state when $\Delta>\Delta_c$ despite the differences in $t_{\perp}$. This insulating state can be understood using the eigenvalues of Dirac point [Eq.\eqref{dirac-point-E}] since they are relevant to the low-energy bands at half filling. We find that when $\Delta\leq{t}_{\perp}$, the system remains metallic as the Fermi level lies between the degenerate eigenvalues $\varepsilon_{1}$ and $\varepsilon_{2}$. In contrast, when $\Delta>t_{\perp}$, $\varepsilon_{-}$ and $\varepsilon_{1}(\varepsilon_{2})$ are inverted compared with the disordered case [ see Fig.~\ref{Structure-bands}(b) and~\ref{gap-band-field}(c) ], resulting in a band-inverted charge-ordered insulating state with a gap of $E_g=\Delta-t_{\perp}$. The latter may be relevant to the zero-field insulating state observed in BBG. For example, $t_{\perp}\approx0.22$eV~\cite{PhysRevLett.99.216802} in BBG, a weakly intralayer CD with critical $\delta{n}_{BA}\approx0.07$ can make it an insulator (Fig.~\ref{gap-n-delta}). Since a reduced threefold symmetry characteristic is observed by high-resolution scanning tunneling microscopy~\cite{stolyarova2007high}, implying an intralayer CD, we argue that the insulating state of BBG at zero field is probably due to the presence of this band-inverted intralayer charge-ordered state.

\begin{figure}[htbp]
\includegraphics[width=0.49\textwidth,height=0.24\textwidth]{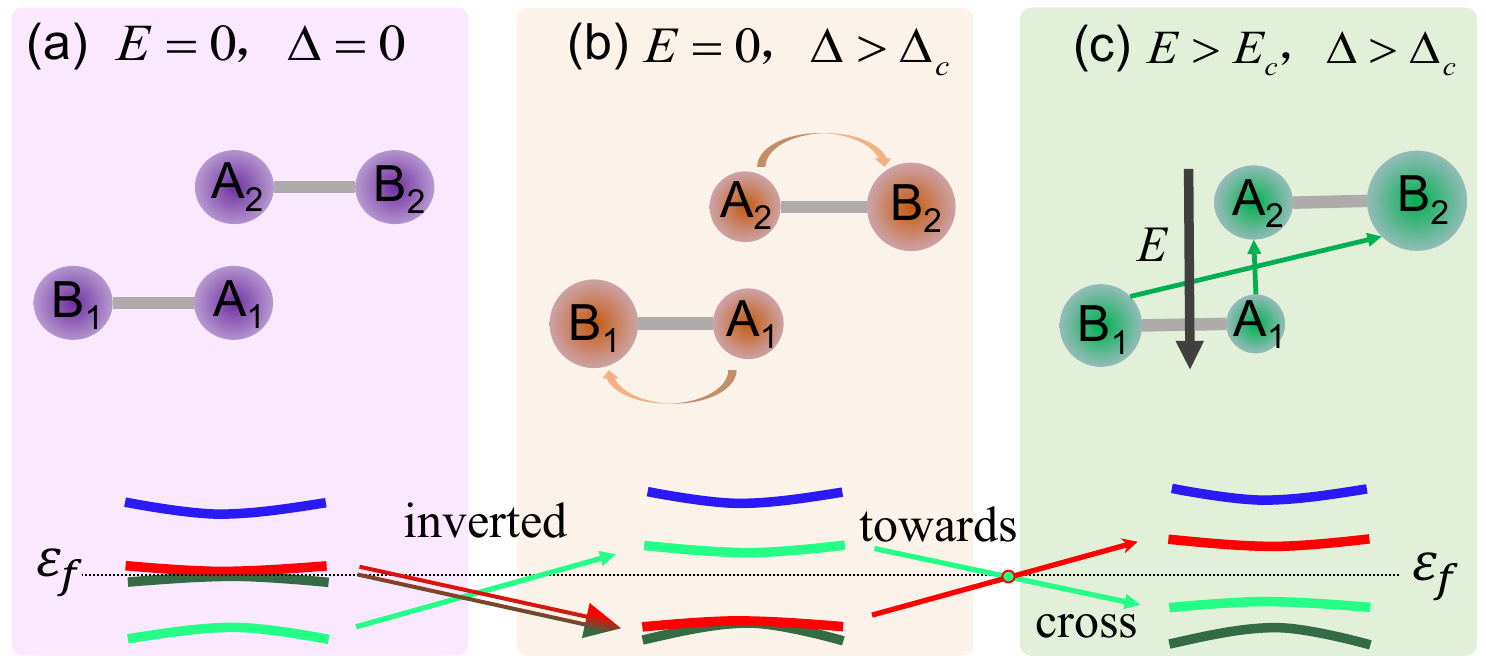}
\caption{The charge distribution and bands in the vicinity of the Dirac point for different cases with (a) absence of electric field and staggered potential, (b) staggered potential is larger than the critical value but without an electric field, (c) both electric field and staggered potential are larger than the critical value. The size of the ball represents the charge population in corresponding site.}
\label{schematic-diagram}
\end{figure}

Proceeding to analyze the effect of an electric field $E$ on the band-inverted charge-ordered insulating state, we find that the gap decreases for $E<E_{c}$, increases for $E>E_{c}$, with the gap closing at critical value $E=E_{c}$, which is reminiscent of the intriguing phenomenon observed experimentally that resistance varies non-monotonically with an electric field~\cite{weitz2010broken}. A downward electric field can drive the electrons from the bottom to the top layers, resulting in interlayer CD. Figure~\ref{gap-band-field}(a) shows $\delta{n}_{21}$ and $\delta{n}_{BA}(E)-\delta{n}_{BA}(0)$ as functions of the electric field, where $\delta{n}_{21}=n_{A_2}$+$n_{B_2}$-$n_{A_1}$-$n_{B_1}$ is the order parameter of the interlayer CD while $\delta{n}_{BA}(E)$ and $\delta{n}_{BA}(0)\thickapprox5.84512\times10^{-3}$ are the order parameters of the field-dependent and zero-field intralayer CDs, respectively. Noticeably, as the band gap is relatively small, we have chosen $t_{\perp}$ and $\Delta$ comparable to the bandgap to clearly demonstrate the detailed evolution of all four relevant low-energy bands with the small electric field, which does not alter the underlying physics. As can be seen, distinct behaviors of $\delta{n}_{21}$ and $\delta{n}_{BA}(E)-\delta{n}_{BA}(0)$ in $E<E_c$ and $E>E_c$ imply a phase transition from phase \uppercase\expandafter{\romannumeral1} to \uppercase\expandafter{\romannumeral2} at $E=E_c$. As a small electric field mainly affects the low-energy bands of the system, the phase transition can also be understood using the eigenvalues of the Dirac point. Figure~\ref{gap-band-field}(b) shows the eigenvalues of the Dirac point varying with the electric field. We find $\varepsilon_2$ and $\varepsilon_{+}$ raise up, whereas $\varepsilon_1$ and $\varepsilon_{-}$ go down with the increase of the electric field. This is because $\varepsilon_2$, $\varepsilon_{+}$, $\varepsilon_1$, and $\varepsilon_{-}$ are mainly contributed by $p_z$ orbitals of $B_1$, $A_1$, $B_2$, and $A_2$ sites, respectively, where the on-site potentials of $B_1$ and $A_1$ sites increase, whereas those of $B_2$ and $A_2$ sites decrease when an electric field is applied. Given that $\varepsilon_{-}$ is higher than $\varepsilon_{2}$ at $E=0$, $\varepsilon_{-}$ and $\varepsilon_{2}$ will move towards and then cross each other as shown in Fig.~\ref{gap-band-field}(c) to~\ref{gap-band-field}(f) [or in Fig.~\ref{schematic-diagram}(b) and~\ref{schematic-diagram}(c)]. Consequently, the gap decreases for $E<E_{c}$ and increases for $E>E_{c}$. Notably, the gap closes at the critical electric field $E_{c}=(\Delta^{2}-t_{\perp}^{2})/(ed\Delta)$. Thus, the gap of the band-inverted intralayer charge-ordered insulating state varies non-monotonously with the electric field.

In brief, the findings above can be summarized as a schematic illustration in Fig.~\ref{schematic-diagram}, demonstrating the following physics. (\romannumeral1) An intralayer charge-ordered state can open an intrinsic gap in the BBG lattice at zero field when band inversion between the two touched bands and the lower band of the two untouched bands occurs, compared with the disordered case [see Fig.~\ref{schematic-diagram}(b) and~\ref{schematic-diagram}(a)]. (\romannumeral2) When an electric field is applied to this charge-ordered insulating state, the upper band of the two touched bands and the lower band of the two untouched bands move towards and then cross each other as shown from Fig.~\ref{schematic-diagram}(b) to~\ref{schematic-diagram}(c), resulting in a non-monotonic behavior of the gap that closes and then reopens with the electric field. As we proposed in Section~\ref{model-and-method} that the interlayer VdW interactions can effectively generate a staggered potential between intralayer inequivalent sublattices. Besides, the ripples which naturally exist in BBG may favor a charge-ordered state. Thus, this band-inverted intralayer charge-ordered insulating state may be the key to the non-monotonic resistance phenomenon observed in BBG, and it is imperative to study the effect of previously ignored interlayer VdW interactions and ripples.
\begin{figure}[htbp]
\includegraphics[width=0.44\textwidth,height=0.38\textwidth]{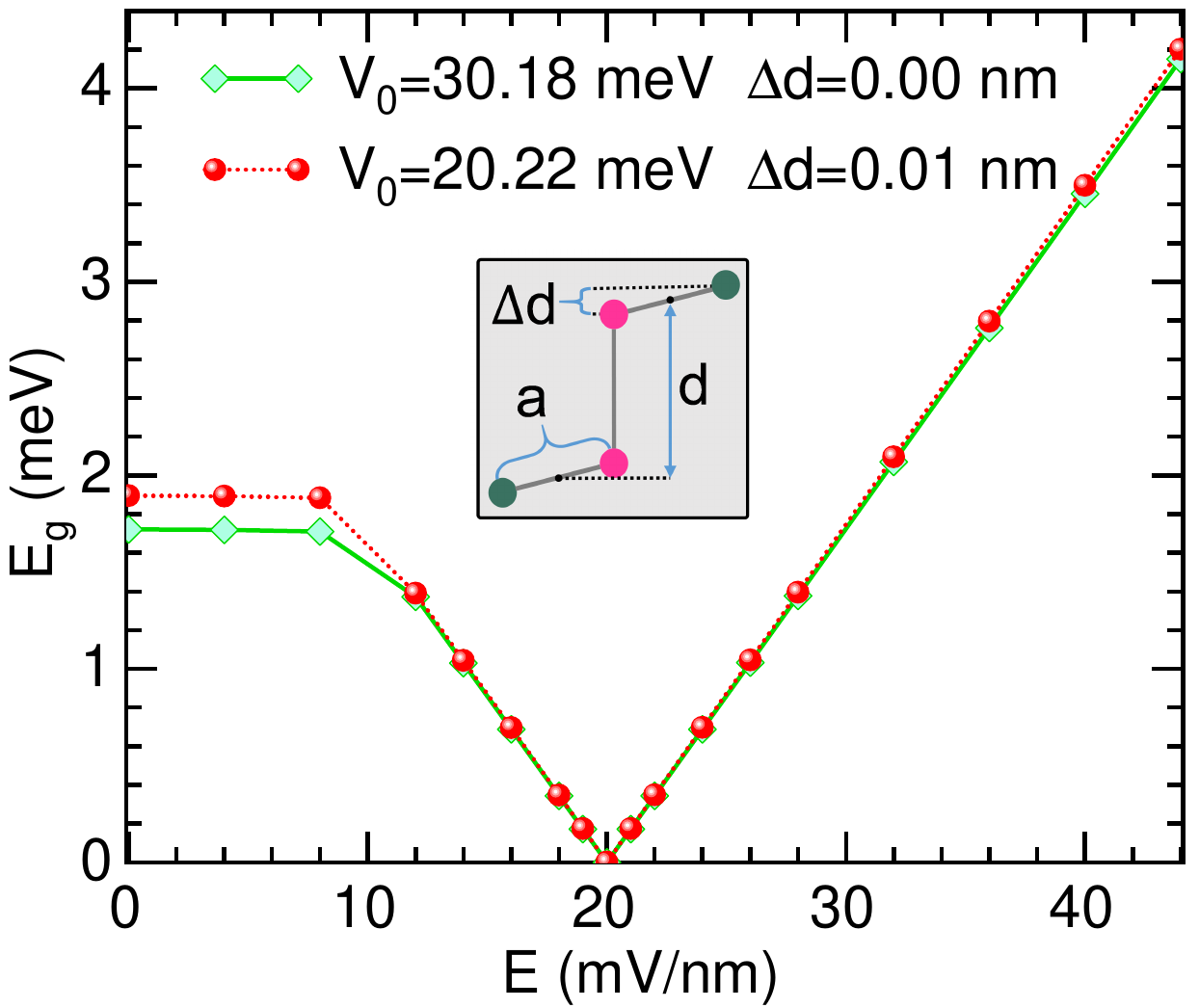}
\caption{The gaps of flat and rippled structures as functions of the electric field. $d=0.34$ nm and $a=0.142$ nm are used. $\Delta{d}=0.01$ nm in rippled structure which is presented in inset.}
\label{gap-DFT}
\end{figure}

To validate our proposal that the intriguing phenomenon observed in BBG is due to the presence of the aforementioned band-inverted charge-ordered state, the model presented in Section~\ref{model-and-method}, specifically Eq.\eqref{DFT+vdw+E}, which includes the effect of interlayer VdW interactions and ripples, is employed to calculate the gap of BBG. For a given strength of interlayer VdW interactions $V_0$, the effective Hamiltonian $H_{eff}$, especially the staggered potential $\Delta$, should be self-consistently determined. Thus, we derive firstly the potential energy difference between A and B sublattices using the method presented in Section~\ref{model-and-method}. We find $\delta{U}_{AB}\approx1.08NV_0$ and $\delta{U}_{AB}\approx2.08NV_0$ for flat structure and Peierls-type rippled structure with $\Delta{d}=0.01$nm, respectively, where $N$ is the number of unit cells. Using Eq.\eqref{eqn:energydiff}, the effective staggered potential is calculated as $\Delta=\delta{U}_{AB}/(N\delta{n}_{BA})$. Thus, $\delta{n}_{BA}$ and $\Delta$ are self-consistently determined by combining this equation with Eq.\eqref{DFT+vdw+E} as long as $V_0$ is given, indicating that $H_{eff}$ can be determined self-consistently. Therefore, the gaps and CDs of bilayer graphene for a given strength of interlayer VdW interactions are calculated. Figure~\ref{gap-DFT} shows the gaps of BBG as functions of the electric field for flat and rippled structures with the strength of interlayer VdW interactions of $V_0=30.18$ meV and $V_0=20.22$ meV, respectively, where the gap decreases for $E<E_{c}$, increases for $E>E_{c}$, with the closure of the gap at $E_{c}=20$ mV/nm. As the resistance $R\propto{exp[E_g/(k_BT)]}$, our results are in excellent agreement with the experimental phenomenon that the resistance decreases and then increases with the electric field, where minimal resistance is at critical electric field $E_c\approx20$ mV/nm~\cite{weitz2010broken} and the zero-field gap is $E_g\approx2$ meV~\cite{velasco2012transport,bao2012evidence,freitag2012spontaneously,veligura2012transport}. It is necessary to mention that a weakly spontaneous charge-ordered state occurs for both cases with intralayer CD $\delta{n}_{BA}=0.10\sim0.12$, where $\delta{n}_{BA}$ changes very little with the electric field  while $\delta{n}_{21}$ increases from 0 to $3\times10^{-4}$. In addition, a smaller $V_0$ can lead to the same behavior of gap for the rippled structure compared with the flat case, suggesting that the ripple concerned and interlayer VdW interactions are cooperative. Thus, interlayer VdW interactions cooperating with ripples can effectively generate a staggered potential between inequivalent sublattices, which induces an intralayer charge-ordered insulating state, resulting in the experimental phenomenon observed in BBG. The strength of the interlayer VdW interactions is of the order of $10$ meV. Please note that $R_{ss^{\prime}}$s are not tunable parameters which are determined by the force equilibrium condition. The critical values of $V_0$ that causes the metal-to-insulator phase transition for flat and ripple structures are 29.85 meV and 20.00 meV, respectively.

\section{discussion}\label{BLG:discussion}
\begin{figure}[htbp]
\subfigure{\includegraphics[width=0.48\textwidth,height=0.24\textwidth]{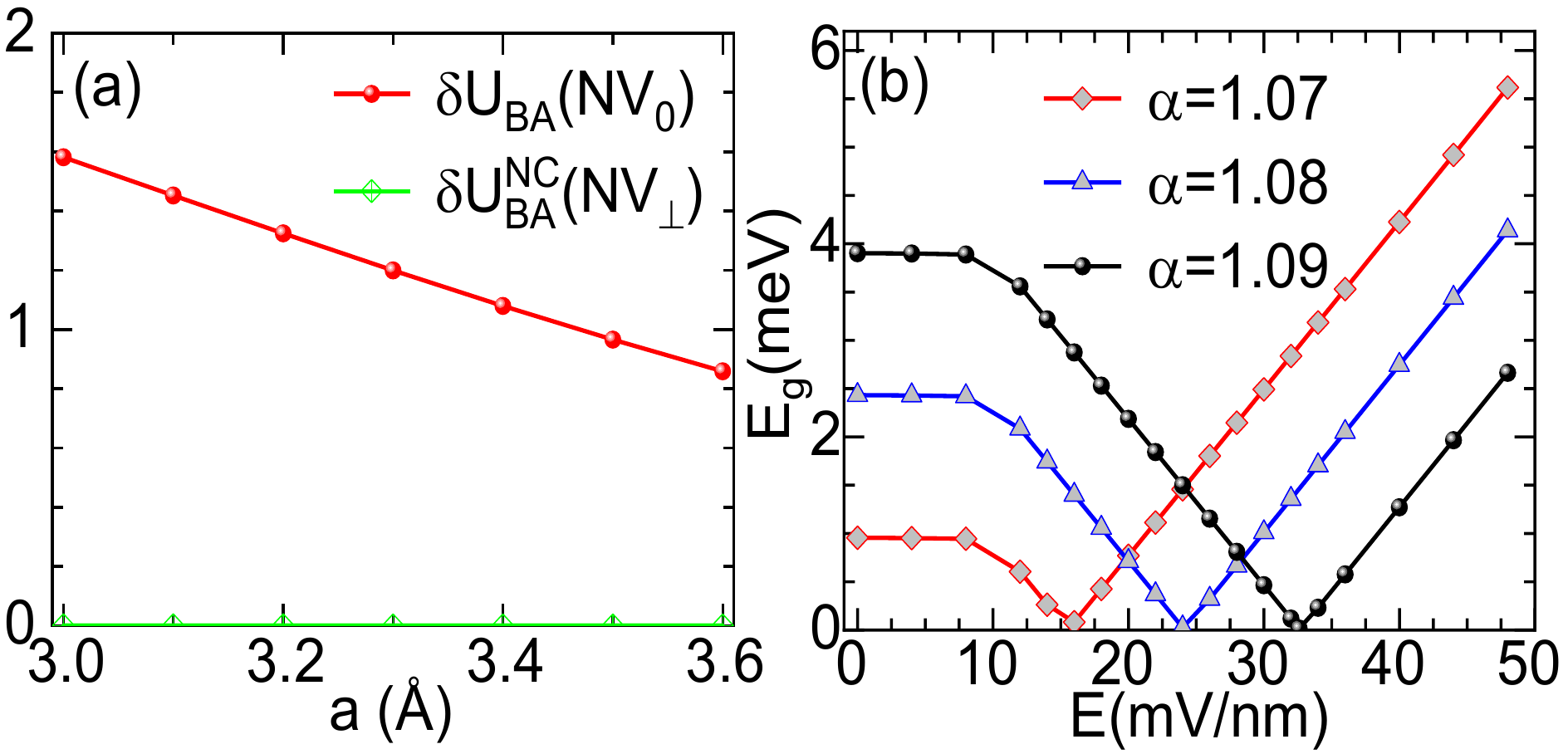}}
\caption{(a) The potential energy differences between inequivalent sublattices generated individually by the interlayer Coulomb interactions ($\delta{U}_{BA}^{NC}$) and that generated individually by the interlayer VdW interactions ($\delta{U}_{BA}$) for the flat structure as functions of interlayer distance $d$, where $V_0$ is the strength of the interlayer VdW interactions and $V_{\perp}$ is the strength of the nearest neighbor interlayer Coulomb interaction. $V(r)=V_{\perp}\frac{d}{\boldsymbol{r}}$ is used for the interlayer Coulomb interactions. $N$ is the number of unit cells. (b) The calculated band gap of model \eqref{DFT+vdw+E} as functions of the electric field, where the Kolmogorov-Crespi potential is used to derive $H_{eff}$ and $\alpha$ is the correction factor for the repulsive term of the Kolmogorov-Crespi potential in BBG as introduced in Appendix \ref{KCP}.}
\label{E-difference}
\end{figure}
Here, a simple model has been used to demonstrate the underlying physics for the intriguing phenomenon observed in BBG that the resistance decreases and then increases with the electric field at low temperatures. We ascribe this phenomenon to the presence of a band-inverted charge-ordered insulating state in BBG. Our proposal is further confirmed by combining DFT calculations with model calculations, where we take into account the effect of the interlayer VdW interactions and ripples. Our results are reliable because our calculations include not only the effects of the nonlocal Coulomb interactions and remote hoppings at the DFT level but also the previously ignored ingredients, namely interlayer VdW interactions and ripple in the layer. Noticeably, although the interlayer Coulomb interactions are stronger than the interlayer VdW interactions, the potential difference between inequivalent sublattices generated by the interlayer Coulomb interactions is much smaller than that generated by the interlayer VdW interactions as presented in Fig.~\ref{E-difference}(a). Thus, the interlayer VdW interactions play a major role in determining the intralayer charge-ordered state of BBG. However, this key ingredient is often ignored by previous studies.

The validity of modeling BBG by a bilayer honeycomb lattice with staggered potentials is as follows: Although DFT can provide a reasonable energy difference between AA-stacking bilayer graphene and BBG due to cancellation of the uncertainty from interlayer interactions simultaneously\cite{kolmogorov2005registry}, it fails to capture the interlayer interactions for each structure individually. Therefore, it is necessary to model the uncertainty arising from interlayer interactions, which may induce a staggered potential between $A_{1(2)}$ and $B_{1(2)}$ sublattices due to their inequivalent interlayer environment. Indeed, our results suggest the presence of a staggered potential due to the interlayer interactions, e.g., VdW interactions. The calculated small band gap, which is consistent with the experimental observations in bilayer graphene devices\cite{velasco2012transport,bao2012evidence,freitag2012spontaneously,veligura2012transport}, strongly indicates that the system is in the critical region of insulator-to-metal transition.

Although varieties of corrections such as
vdW-DF~\cite{dion2004van},
TS-vdW~\cite{tkatchenko2009accurate},
vdW-DF-C~\cite{cooper2010van},
DFT-D~\cite{grimme2010consistent} and so on have been proposed, the electronic properties of the BBG and graphite obtained from these corrections are diverse from each other~\cite{lebedeva2011interlayer,del2019layer}.  Thus, in order to gain insight into
the effect of VdW interactions and eliminate the uncertainty from different choices of corrections, it is necessary to treat the interlayer VdW interactions as free parameters, namely modeling the effect of interlayer VdW interactions. In our calculations, the model Hamiltonian for interlayer VdW interactions is obtained naturally from the interatomic Lennard–Jones potential, where the strength of interlayer VdW interactions $V_0$ serves as the sole free parameter except for the electric field in Eq.\eqref{DFT+vdw+E}. Noticeably, the isotropic nature of the Lennard–Jones VdW potential can not capture the anisotropic properties of BBG. Thus, we also employ the Kolmogorov-Crespi potential\cite{kolmogorov2005registry} to take into account the interlayer interactions with anisotropy as introduced in Appendix \ref{KCP}. Similar behavior that the band gap varies non-monotonically with the electric field has also been observed as shown in Fig.\ref{E-difference}(b).

Although several correlated symmetry-broken gapped states with parabolic dispersion relation have been proposed at zero field, namely LAF, CAF, QAH, and QSH, the low-energy bands of BBG observed experimentally at a small applied magnetic field cannot be explained within the framework of parabolic bands, which predicts roughly equidistant Landau levels at low temperatures~\cite{mayorov2011interaction}. Besides, there is a pronounced asymmetry in the cyclotron mass between hole- and electron-doping~\cite{PhysRevLett.99.216802}. These findings raise doubts about the candidates which exhibit a parabolic dispersion relation with particle-hole symmetry near the Fermi level. Moreover, as the temperature increases from zero, two resistance transitions occur. One transition is observed at $\sim12$K~\cite{nam2018family}, while the other occurs at $200\sim250$K~\cite{liu2017observation}, which is suggested to be caused by the interlayer ripple scattering effect. As the charge-ordered state we proposed still exists even when the gap is closed, it may suggest that the former transition corresponds to the evolution from the charge-ordered insulating state to the charge-ordered metallic state, while the latter transition corresponds to the change from the charge-ordered metallic state to the disordered state.


Ripples are inherent features of BBG, arising from the natural undulations of suspended graphene sheets. It has been proposed that suspended graphene sheets are not perfectly flat showing ripples with an amplitude of about 1 nm~\cite{stolyarova2007high,meyer2007structure,fasolino2007intrinsic} with dislocations~\cite{butz2014dislocations}. Here, for simplicity, we take the Peierls-type ripple into account, which is energetically favored by elastic effects~\cite{PhysRevB.79.045421}. However, it should be noted that the ripples in BBG exhibit a complex nature. Thus, it is interesting to study how the experimentally observed ripples cooperate with interlayer VdW interactions to affect the properties of the intralayer charge-ordered state.

Although the charge-ordered state we studied has been previously investigated~\cite{dahal2010charge}, the properties of this charge-ordered state under an external field have not been explored before. Noticeably, a low-energy theory based on a $2\times2$ Hamiltonian matrix with consideration of the charge-ordered characteristic is used to study the properties of BBG~\cite{nandkishore2010quantum}. However, it fails to deal with the properties of the charge-ordered state concerned here, where the low-energy bands are inverted. Besides, although low-energy theory based on a $4\times4$ Hamiltonian matrix is also proposed, it does not take into account the effect of a charge-ordered state~\cite{PhysRevB.74.161403,nilsson2006electron}.

Here, we study the phenomenon observed in BBG that the gap varies non-monotonously with the electric field. Our results strongly suggest that the ground state of BBG is a charge-ordered insulating state. Therefore, further experiments are needed to confirm the ground state of BBG. There are two experimental approaches to identify this state: one is angle-resolved photoelectron spectroscopy and the other is scanning tunneling spectroscopy. An experiment based on angular-resolved photoemission spectroscopy should be done at low temperatures, without external perturbations, to detect the low-energy bands of BBG. If the low-energy bands are inverted, the ground state of BBG is a charge-ordered state. Alternatively, the scanning tunneling spectroscopy would be sensitive to the charge ordering at atomic scale, allowing one to measure spatial variations of the local density of states to determine the ground state of BBG.
\section{conclusion}\label{BLG:conclusion}
In conclusion, an intriguing phenomenon that the resistance varies non-monotonously with an electric field applied perpendicular to the plane has been observed at low temperatures in BBG. Here, we suggest that this phenomenon is probably due to the presence of a spontaneous charge-ordered insulating state in BBG. The underlying physics is illustrated by a simple model on BBG lattice with staggered potential between inequivalent sublattices. To validate our proposal, we combine DFT calculations with model calculations to include the effect of the interlayer VdW interactions and ripples. We find that the interlayer VdW interactions cooperating with ripples can effectively generate a staggered potential in BBG. Remarkably, we have successfully reproduced the gap amplitude and the critical electric field when the strength of the interlayer VdW interactions is on the order of $10$ meV. Our results provide a new perspective on the non-monotonic resistance phenomenon in BBG and suggest that the ground state of BBG is likely a charge-ordered state. We argue that angular-resolved photoemission spectroscopy studies at zero field or scanning tunneling spectroscopy can be used to identify the ground state at low temperatures.
\begin{center}
\textbf{ACKNOWLEDGEMENT}
\end{center}
This work is supported by National Natural Science Foundation of China (Grants No. 12274324, No. 12004283) and Shanghai Science and Technology Commission (Grant No. 21JC405700).
\appendix
\section{Wannier fittings to the low-energy DFT bands}\label{DFT-fit}
\begin{figure}[htbp]
\includegraphics[width=0.48\textwidth,height=0.23\textwidth]{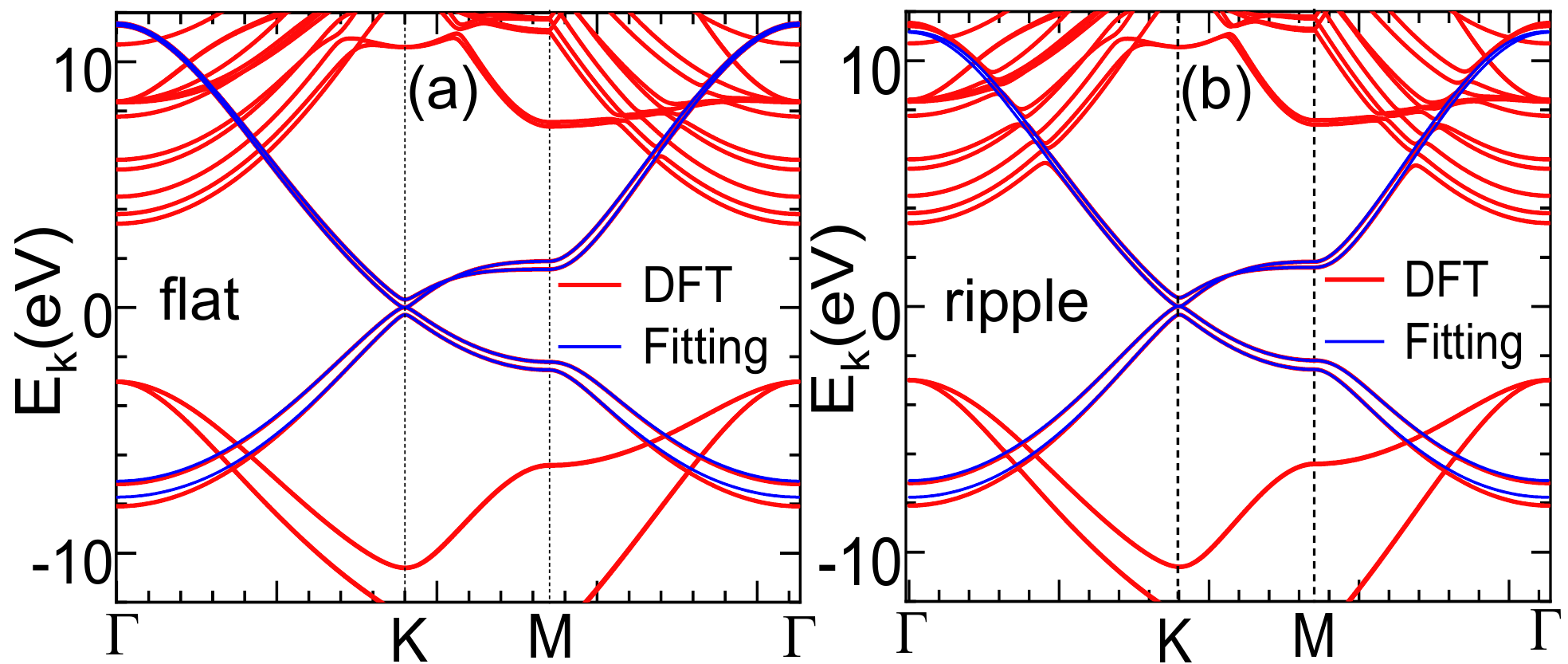}
\caption{The bands obtained from DFT calculations and the corresponding wannier fittings for flat structure (a) and for ripple structure with $\Delta{d}=0.01$nm (b).}
\label{wannierfit}
\end{figure}
Due to the fact that only including $t_0$ and $t_\perp$ is not sufficient to well describe the low-energy DFT bands of BBG, we establish a tight binding model $H_D$ with long-range hoppings to accurately describe the entire dispersions of low-energy DFT bands. For both flat and ripple structures, the fitted bands provided by the tight-binding Hamiltonian of all $p_z$ orbitals (blue) and DFT bands (red) match well as shown in Fig.\ref{wannierfit}.
\section{The total potential energy difference between A-type and B-type sublattices derived from the Kolmogorov-Crespi potential}\label{KCP}
It has been pointed out that the isotropic nature of the Lennard–Jones VdW potential can not capture the anisotropic properties of the graphitic systems. Then, Kolmogorov and Crespi take into account the in-plane and out-of-plane anisotropy, proposing the so-called Kolmogorov-Crespi potential to describe the interlayer interactions in graphite systems\cite{kolmogorov2005registry}. The interatomic Kolmogorov-Crespi potential depicted the graphitic systems reads
\begin{small}
\begin{equation}
w_{ij}^{ss^{\prime}}=e^{-\lambda(r_{is}^{js^{\prime}}-d)}\Big[C+f(\rho_{is}^{js^{\prime}})+f(\rho_{js^{\prime}}^{is})\Big]
-A\bigg(\frac{r_{is}^{js^{\prime}}}{d}\bigg)^6,
\end{equation}
\end{small}
with
\begin{equation}
\begin{aligned}
r_{is}^{js^{\prime}}&=|\boldsymbol{r_{is}}-\boldsymbol{r_{js^{\prime}}}|,\\
(\rho_{is}^{js^{\prime}})^2&=(r_{is}^{js^{\prime}})^2-(\boldsymbol{n_{is}}\cdot\boldsymbol{r}_{is}^{js^{\prime}})^2,\\
(\rho_{js^{\prime}}^{is})^2&=(r_{js^{\prime}}^{is})^2-(\boldsymbol{n_{js^{\prime}}}\cdot\boldsymbol{r}_{js^{\prime}}^{is})^2,\\
f(\rho_{is}^{js^{\prime}})&=e^{-\rho_{is}^{js^{\prime}}/\delta}\sum\limits_{n=0}^{2}{C_{2n}}(\rho_{is}^{js^{\prime}}/\delta)^{2n},
\end{aligned}
\end{equation}
where $d$ is the interlayer distance while the other constants are as follows:
\begin{equation}
\begin{aligned}
C_0&=15.71~\text{meV}\quad C_2=12.29~\text{meV}\\
C_4&=4.933~\text{meV}\quad C=3.030~\text{meV}\\
\delta&=0.578~\text{{\AA}} \quad \lambda=3.629~\text{{\AA}}^{-1} \quad A=10.238~\text{meV}
\end{aligned}
\end{equation}
Besides, it was shown that for a layered system composed of two monolayer planes like BBG, the Casimir force plays a significant role in phase transitions\cite{flachi2013strongly}. As the Casimir force generates an additional attractive interaction between the two planes, the repulsive term of the Kolmogorov-Crespi potential for BBG has to be larger than that of graphite. Thus, the interatomic Kolmogorov-Crespi potential describing BBG can be written as
\begin{small}
\begin{equation}
w_{ij}^{ss^{\prime}}=\alpha e^{-\lambda(r_{is}^{js^{\prime}}-d)}\Big[C+f(\rho_{is}^{js^{\prime}})+f(\rho_{js^{\prime}}^{is})\Big]
-A\bigg(\frac{r_{is}^{js^{\prime}}}{d}\bigg)^6,
\end{equation}
\end{small}
where $\alpha$ is the correction factor for the repulsive term of the Kolmogorov-Crespi potential in BBG which should be slightly larger than 1 and the other constants remain the same as those of graphite. Substituting this equation into Eq.\eqref{eqn:VdW} can determine the staggered potentials and consequently calculate the band gap of BBG once $\alpha$ is given by applying equations of \eqref{DFT+vdw+E}, \eqref{deltaU}, and \eqref{eqn:energydiff} subsequently. The calculated band gap of BBG under the applied electric field is shown in Fig.\ref{E-difference}(b), which qualitatively agrees with the result shown in Fig.\ref{gap-DFT} where the Lennard–Jones VdW potential is used.

\bibliography{BLG_reference}

\providecommand{\noopsort}[1]{}\providecommand{\singleletter}[1]{#1}%
\begin{thebibliography}{72}%
\makeatletter
\providecommand \@ifxundefined [1]{%
 \@ifx{#1\undefined}
}%
\providecommand \@ifnum [1]{%
 \ifnum #1\expandafter \@firstoftwo
 \else \expandafter \@secondoftwo
 \fi
}%
\providecommand \@ifx [1]{%
 \ifx #1\expandafter \@firstoftwo
 \else \expandafter \@secondoftwo
 \fi
}%
\providecommand \natexlab [1]{#1}%
\providecommand \enquote  [1]{``#1''}%
\providecommand \bibnamefont  [1]{#1}%
\providecommand \bibfnamefont [1]{#1}%
\providecommand \citenamefont [1]{#1}%
\providecommand \href@noop [0]{\@secondoftwo}%
\providecommand \href [0]{\begingroup \@sanitize@url \@href}%
\providecommand \@href[1]{\@@startlink{#1}\@@href}%
\providecommand \@@href[1]{\endgroup#1\@@endlink}%
\providecommand \@sanitize@url [0]{\catcode `\\12\catcode `\$12\catcode
  `\&12\catcode `\#12\catcode `\^12\catcode `\_12\catcode `\%12\relax}%
\providecommand \@@startlink[1]{}%
\providecommand \@@endlink[0]{}%
\providecommand \url  [0]{\begingroup\@sanitize@url \@url }%
\providecommand \@url [1]{\endgroup\@href {#1}{\urlprefix }}%
\providecommand \urlprefix  [0]{URL }%
\providecommand \Eprint [0]{\href }%
\providecommand \doibase [0]{http://dx.doi.org/}%
\providecommand \selectlanguage [0]{\@gobble}%
\providecommand \bibinfo  [0]{\@secondoftwo}%
\providecommand \bibfield  [0]{\@secondoftwo}%
\providecommand \translation [1]{[#1]}%
\providecommand \BibitemOpen [0]{}%
\providecommand \bibitemStop [0]{}%
\providecommand \bibitemNoStop [0]{.\EOS\space}%
\providecommand \EOS [0]{\spacefactor3000\relax}%
\providecommand \BibitemShut  [1]{\csname bibitem#1\endcsname}%
\let\auto@bib@innerbib\@empty
\bibitem [{\citenamefont {Novoselov}\ \emph {et~al.}(2006)\citenamefont
  {Novoselov}, \citenamefont {McCann}, \citenamefont {Morozov}, \citenamefont
  {Fal’ko}, \citenamefont {Katsnelson}, \citenamefont {Zeitler},
  \citenamefont {Jiang}, \citenamefont {Schedin},\ and\ \citenamefont
  {Geim}}]{novoselov2006unconventional}%
  \BibitemOpen
  \bibfield  {author} {\bibinfo {author} {\bibfnamefont {Kostya~S}\
  \bibnamefont {Novoselov}}, \bibinfo {author} {\bibfnamefont {Edward}\
  \bibnamefont {McCann}}, \bibinfo {author} {\bibfnamefont {SV}~\bibnamefont
  {Morozov}}, \bibinfo {author} {\bibfnamefont {Vladimir~I}\ \bibnamefont
  {Fal’ko}}, \bibinfo {author} {\bibfnamefont {MI}~\bibnamefont
  {Katsnelson}}, \bibinfo {author} {\bibfnamefont {U}~\bibnamefont {Zeitler}},
  \bibinfo {author} {\bibfnamefont {D}~\bibnamefont {Jiang}}, \bibinfo {author}
  {\bibfnamefont {F}~\bibnamefont {Schedin}}, \ and\ \bibinfo {author}
  {\bibfnamefont {AK}~\bibnamefont {Geim}},\ }\bibfield  {title} {\enquote
  {\bibinfo {title} {Unconventional quantum hall effect and berry’s phase of
  2$\pi$ in bilayer graphene},}\ }\href@noop {} {\bibfield  {journal} {\bibinfo
   {journal} {Nature physics}\ }\textbf {\bibinfo {volume} {2}},\ \bibinfo
  {pages} {177--180} (\bibinfo {year} {2006})}\BibitemShut {NoStop}%
\bibitem [{\citenamefont {Kou}\ \emph {et~al.}(2014)\citenamefont {Kou},
  \citenamefont {Feldman}, \citenamefont {Levin}, \citenamefont {Halperin},
  \citenamefont {Watanabe}, \citenamefont {Taniguchi},\ and\ \citenamefont
  {Yacoby}}]{kou2014electron}%
  \BibitemOpen
  \bibfield  {author} {\bibinfo {author} {\bibfnamefont {Angela}\ \bibnamefont
  {Kou}}, \bibinfo {author} {\bibfnamefont {Benjamin~E}\ \bibnamefont
  {Feldman}}, \bibinfo {author} {\bibfnamefont {Andrei~J}\ \bibnamefont
  {Levin}}, \bibinfo {author} {\bibfnamefont {Bertrand~I}\ \bibnamefont
  {Halperin}}, \bibinfo {author} {\bibfnamefont {Kenji}\ \bibnamefont
  {Watanabe}}, \bibinfo {author} {\bibfnamefont {Takashi}\ \bibnamefont
  {Taniguchi}}, \ and\ \bibinfo {author} {\bibfnamefont {Amir}\ \bibnamefont
  {Yacoby}},\ }\bibfield  {title} {\enquote {\bibinfo {title} {Electron-hole
  asymmetric integer and fractional quantum hall effect in bilayer graphene},}\
  }\href@noop {} {\bibfield  {journal} {\bibinfo  {journal} {Science}\ }\textbf
  {\bibinfo {volume} {345}},\ \bibinfo {pages} {55--57} (\bibinfo {year}
  {2014})}\BibitemShut {NoStop}%
\bibitem [{\citenamefont {Cao}\ \emph {et~al.}(2018{\natexlab{a}})\citenamefont
  {Cao}, \citenamefont {Fatemi}, \citenamefont {Fang}, \citenamefont
  {Watanabe}, \citenamefont {Taniguchi}, \citenamefont {Kaxiras},\ and\
  \citenamefont {Jarillo-Herrero}}]{cao2018unconventional}%
  \BibitemOpen
  \bibfield  {author} {\bibinfo {author} {\bibfnamefont {Yuan}\ \bibnamefont
  {Cao}}, \bibinfo {author} {\bibfnamefont {Valla}\ \bibnamefont {Fatemi}},
  \bibinfo {author} {\bibfnamefont {Shiang}\ \bibnamefont {Fang}}, \bibinfo
  {author} {\bibfnamefont {Kenji}\ \bibnamefont {Watanabe}}, \bibinfo {author}
  {\bibfnamefont {Takashi}\ \bibnamefont {Taniguchi}}, \bibinfo {author}
  {\bibfnamefont {Efthimios}\ \bibnamefont {Kaxiras}}, \ and\ \bibinfo {author}
  {\bibfnamefont {Pablo}\ \bibnamefont {Jarillo-Herrero}},\ }\bibfield  {title}
  {\enquote {\bibinfo {title} {Unconventional superconductivity in magic-angle
  graphene superlattices},}\ }\href@noop {} {\bibfield  {journal} {\bibinfo
  {journal} {Nature}\ }\textbf {\bibinfo {volume} {556}},\ \bibinfo {pages}
  {43--50} (\bibinfo {year} {2018}{\natexlab{a}})}\BibitemShut {NoStop}%
\bibitem [{\citenamefont {Lu}\ \emph {et~al.}(2019)\citenamefont {Lu},
  \citenamefont {Stepanov}, \citenamefont {Yang}, \citenamefont {Xie},
  \citenamefont {Aamir}, \citenamefont {Das}, \citenamefont {Urgell},
  \citenamefont {Watanabe}, \citenamefont {Taniguchi}, \citenamefont {Zhang}
  \emph {et~al.}}]{lu2019superconductors}%
  \BibitemOpen
  \bibfield  {author} {\bibinfo {author} {\bibfnamefont {Xiaobo}\ \bibnamefont
  {Lu}}, \bibinfo {author} {\bibfnamefont {Petr}\ \bibnamefont {Stepanov}},
  \bibinfo {author} {\bibfnamefont {Wei}\ \bibnamefont {Yang}}, \bibinfo
  {author} {\bibfnamefont {Ming}\ \bibnamefont {Xie}}, \bibinfo {author}
  {\bibfnamefont {Mohammed~Ali}\ \bibnamefont {Aamir}}, \bibinfo {author}
  {\bibfnamefont {Ipsita}\ \bibnamefont {Das}}, \bibinfo {author}
  {\bibfnamefont {Carles}\ \bibnamefont {Urgell}}, \bibinfo {author}
  {\bibfnamefont {Kenji}\ \bibnamefont {Watanabe}}, \bibinfo {author}
  {\bibfnamefont {Takashi}\ \bibnamefont {Taniguchi}}, \bibinfo {author}
  {\bibfnamefont {Guangyu}\ \bibnamefont {Zhang}},  \emph {et~al.},\ }\bibfield
   {title} {\enquote {\bibinfo {title} {Superconductors, orbital magnets and
  correlated states in magic-angle bilayer graphene},}\ }\href@noop {}
  {\bibfield  {journal} {\bibinfo  {journal} {Nature}\ }\textbf {\bibinfo
  {volume} {574}},\ \bibinfo {pages} {653--657} (\bibinfo {year}
  {2019})}\BibitemShut {NoStop}%
\bibitem [{\citenamefont {Park}\ \emph {et~al.}(2019)\citenamefont {Park},
  \citenamefont {Kim}, \citenamefont {Cho},\ and\ \citenamefont
  {Lee}}]{park2019higher}%
  \BibitemOpen
  \bibfield  {author} {\bibinfo {author} {\bibfnamefont {Moon~Jip}\
  \bibnamefont {Park}}, \bibinfo {author} {\bibfnamefont {Youngkuk}\
  \bibnamefont {Kim}}, \bibinfo {author} {\bibfnamefont {Gil~Young}\
  \bibnamefont {Cho}}, \ and\ \bibinfo {author} {\bibfnamefont {SungBin}\
  \bibnamefont {Lee}},\ }\bibfield  {title} {\enquote {\bibinfo {title}
  {Higher-order topological insulator in twisted bilayer graphene},}\
  }\href@noop {} {\bibfield  {journal} {\bibinfo  {journal} {Physical review
  letters}\ }\textbf {\bibinfo {volume} {123}},\ \bibinfo {pages} {216803}
  (\bibinfo {year} {2019})}\BibitemShut {NoStop}%
\bibitem [{\citenamefont {Park}\ and\ \citenamefont
  {Louie}(2010)}]{park2010tunable}%
  \BibitemOpen
  \bibfield  {author} {\bibinfo {author} {\bibfnamefont {Cheol-Hwan}\
  \bibnamefont {Park}}\ and\ \bibinfo {author} {\bibfnamefont {Steven~G}\
  \bibnamefont {Louie}},\ }\bibfield  {title} {\enquote {\bibinfo {title}
  {Tunable excitons in biased bilayer graphene},}\ }\href@noop {} {\bibfield
  {journal} {\bibinfo  {journal} {Nano letters}\ }\textbf {\bibinfo {volume}
  {10}},\ \bibinfo {pages} {426--431} (\bibinfo {year} {2010})}\BibitemShut
  {NoStop}%
\bibitem [{\citenamefont {Ju}\ \emph {et~al.}(2017)\citenamefont {Ju},
  \citenamefont {Wang}, \citenamefont {Cao}, \citenamefont {Taniguchi},
  \citenamefont {Watanabe}, \citenamefont {Louie}, \citenamefont {Rana},
  \citenamefont {Park}, \citenamefont {Hone}, \citenamefont {Wang} \emph
  {et~al.}}]{ju2017tunable}%
  \BibitemOpen
  \bibfield  {author} {\bibinfo {author} {\bibfnamefont {Long}\ \bibnamefont
  {Ju}}, \bibinfo {author} {\bibfnamefont {Lei}\ \bibnamefont {Wang}}, \bibinfo
  {author} {\bibfnamefont {Ting}\ \bibnamefont {Cao}}, \bibinfo {author}
  {\bibfnamefont {Takashi}\ \bibnamefont {Taniguchi}}, \bibinfo {author}
  {\bibfnamefont {Kenji}\ \bibnamefont {Watanabe}}, \bibinfo {author}
  {\bibfnamefont {Steven~G}\ \bibnamefont {Louie}}, \bibinfo {author}
  {\bibfnamefont {Farhan}\ \bibnamefont {Rana}}, \bibinfo {author}
  {\bibfnamefont {Jiwoong}\ \bibnamefont {Park}}, \bibinfo {author}
  {\bibfnamefont {James}\ \bibnamefont {Hone}}, \bibinfo {author}
  {\bibfnamefont {Feng}\ \bibnamefont {Wang}},  \emph {et~al.},\ }\bibfield
  {title} {\enquote {\bibinfo {title} {Tunable excitons in bilayer graphene},}\
  }\href@noop {} {\bibfield  {journal} {\bibinfo  {journal} {Science}\ }\textbf
  {\bibinfo {volume} {358}},\ \bibinfo {pages} {907--910} (\bibinfo {year}
  {2017})}\BibitemShut {NoStop}%
\bibitem [{\citenamefont {Sui}\ \emph {et~al.}(2015)\citenamefont {Sui},
  \citenamefont {Chen}, \citenamefont {Ma}, \citenamefont {Shan}, \citenamefont
  {Tian}, \citenamefont {Watanabe}, \citenamefont {Taniguchi}, \citenamefont
  {Jin}, \citenamefont {Yao}, \citenamefont {Xiao} \emph
  {et~al.}}]{sui2015gate}%
  \BibitemOpen
  \bibfield  {author} {\bibinfo {author} {\bibfnamefont {Mengqiao}\
  \bibnamefont {Sui}}, \bibinfo {author} {\bibfnamefont {Guorui}\ \bibnamefont
  {Chen}}, \bibinfo {author} {\bibfnamefont {Liguo}\ \bibnamefont {Ma}},
  \bibinfo {author} {\bibfnamefont {Wen-Yu}\ \bibnamefont {Shan}}, \bibinfo
  {author} {\bibfnamefont {Dai}\ \bibnamefont {Tian}}, \bibinfo {author}
  {\bibfnamefont {Kenji}\ \bibnamefont {Watanabe}}, \bibinfo {author}
  {\bibfnamefont {Takashi}\ \bibnamefont {Taniguchi}}, \bibinfo {author}
  {\bibfnamefont {Xiaofeng}\ \bibnamefont {Jin}}, \bibinfo {author}
  {\bibfnamefont {Wang}\ \bibnamefont {Yao}}, \bibinfo {author} {\bibfnamefont
  {Di}~\bibnamefont {Xiao}},  \emph {et~al.},\ }\bibfield  {title} {\enquote
  {\bibinfo {title} {Gate-tunable topological valley transport in bilayer
  graphene},}\ }\href@noop {} {\bibfield  {journal} {\bibinfo  {journal}
  {Nature Physics}\ }\textbf {\bibinfo {volume} {11}},\ \bibinfo {pages}
  {1027--1031} (\bibinfo {year} {2015})}\BibitemShut {NoStop}%
\bibitem [{\citenamefont {Rozhkov}\ \emph {et~al.}(2016)\citenamefont
  {Rozhkov}, \citenamefont {Sboychakov}, \citenamefont {Rakhmanov},\ and\
  \citenamefont {Nori}}]{rozhkov2016electronic}%
  \BibitemOpen
  \bibfield  {author} {\bibinfo {author} {\bibfnamefont
  {Alexandr~Vladimirovich}\ \bibnamefont {Rozhkov}}, \bibinfo {author}
  {\bibfnamefont {AO}~\bibnamefont {Sboychakov}}, \bibinfo {author}
  {\bibfnamefont {AL}~\bibnamefont {Rakhmanov}}, \ and\ \bibinfo {author}
  {\bibfnamefont {Franco}\ \bibnamefont {Nori}},\ }\bibfield  {title} {\enquote
  {\bibinfo {title} {Electronic properties of graphene-based bilayer
  systems},}\ }\href@noop {} {\bibfield  {journal} {\bibinfo  {journal}
  {Physics Reports}\ }\textbf {\bibinfo {volume} {648}},\ \bibinfo {pages}
  {1--104} (\bibinfo {year} {2016})}\BibitemShut {NoStop}%
\bibitem [{\citenamefont {Oostinga}\ \emph {et~al.}(2008)\citenamefont
  {Oostinga}, \citenamefont {Heersche}, \citenamefont {Liu}, \citenamefont
  {Morpurgo},\ and\ \citenamefont {Vandersypen}}]{oostinga2008gate}%
  \BibitemOpen
  \bibfield  {author} {\bibinfo {author} {\bibfnamefont {Jeroen~B}\
  \bibnamefont {Oostinga}}, \bibinfo {author} {\bibfnamefont {Hubert~B}\
  \bibnamefont {Heersche}}, \bibinfo {author} {\bibfnamefont {Xinglan}\
  \bibnamefont {Liu}}, \bibinfo {author} {\bibfnamefont {Alberto~F}\
  \bibnamefont {Morpurgo}}, \ and\ \bibinfo {author} {\bibfnamefont
  {Lieven~MK}\ \bibnamefont {Vandersypen}},\ }\bibfield  {title} {\enquote
  {\bibinfo {title} {Gate-induced insulating state in bilayer graphene
  devices},}\ }\href@noop {} {\bibfield  {journal} {\bibinfo  {journal} {Nature
  materials}\ }\textbf {\bibinfo {volume} {7}},\ \bibinfo {pages} {151--157}
  (\bibinfo {year} {2008})}\BibitemShut {NoStop}%
\bibitem [{\citenamefont {Feldman}\ \emph {et~al.}(2009)\citenamefont
  {Feldman}, \citenamefont {Martin},\ and\ \citenamefont
  {Yacoby}}]{feldman2009broken}%
  \BibitemOpen
  \bibfield  {author} {\bibinfo {author} {\bibfnamefont {Benjamin~E}\
  \bibnamefont {Feldman}}, \bibinfo {author} {\bibfnamefont {Jens}\
  \bibnamefont {Martin}}, \ and\ \bibinfo {author} {\bibfnamefont {Amir}\
  \bibnamefont {Yacoby}},\ }\bibfield  {title} {\enquote {\bibinfo {title}
  {Broken-symmetry states and divergent resistance in suspended bilayer
  graphene},}\ }\href@noop {} {\bibfield  {journal} {\bibinfo  {journal}
  {Nature Physics}\ }\textbf {\bibinfo {volume} {5}},\ \bibinfo {pages}
  {889--893} (\bibinfo {year} {2009})}\BibitemShut {NoStop}%
\bibitem [{\citenamefont {Taychatanapat}\ and\ \citenamefont
  {Jarillo-Herrero}(2010)}]{taychatanapat2010electronic}%
  \BibitemOpen
  \bibfield  {author} {\bibinfo {author} {\bibfnamefont {Thiti}\ \bibnamefont
  {Taychatanapat}}\ and\ \bibinfo {author} {\bibfnamefont {Pablo}\ \bibnamefont
  {Jarillo-Herrero}},\ }\bibfield  {title} {\enquote {\bibinfo {title}
  {Electronic transport in dual-gated bilayer graphene at large displacement
  fields},}\ }\href@noop {} {\bibfield  {journal} {\bibinfo  {journal}
  {Physical review letters}\ }\textbf {\bibinfo {volume} {105}},\ \bibinfo
  {pages} {166601} (\bibinfo {year} {2010})}\BibitemShut {NoStop}%
\bibitem [{\citenamefont {Maher}\ \emph {et~al.}(2013)\citenamefont {Maher},
  \citenamefont {Dean}, \citenamefont {Young}, \citenamefont {Taniguchi},
  \citenamefont {Watanabe}, \citenamefont {Shepard}, \citenamefont {Hone},\
  and\ \citenamefont {Kim}}]{maher2013evidence}%
  \BibitemOpen
  \bibfield  {author} {\bibinfo {author} {\bibfnamefont {Patrick}\ \bibnamefont
  {Maher}}, \bibinfo {author} {\bibfnamefont {Cory~R}\ \bibnamefont {Dean}},
  \bibinfo {author} {\bibfnamefont {Andrea~F}\ \bibnamefont {Young}}, \bibinfo
  {author} {\bibfnamefont {Takashi}\ \bibnamefont {Taniguchi}}, \bibinfo
  {author} {\bibfnamefont {Kenji}\ \bibnamefont {Watanabe}}, \bibinfo {author}
  {\bibfnamefont {Kenneth~L}\ \bibnamefont {Shepard}}, \bibinfo {author}
  {\bibfnamefont {James}\ \bibnamefont {Hone}}, \ and\ \bibinfo {author}
  {\bibfnamefont {Philip}\ \bibnamefont {Kim}},\ }\bibfield  {title} {\enquote
  {\bibinfo {title} {Evidence for a spin phase transition at charge neutrality
  in bilayer graphene},}\ }\href@noop {} {\bibfield  {journal} {\bibinfo
  {journal} {Nature Physics}\ }\textbf {\bibinfo {volume} {9}},\ \bibinfo
  {pages} {154--158} (\bibinfo {year} {2013})}\BibitemShut {NoStop}%
\bibitem [{\citenamefont {Li}\ \emph {et~al.}(2019)\citenamefont {Li},
  \citenamefont {Fu}, \citenamefont {Yin}, \citenamefont {Watanabe},
  \citenamefont {Taniguchi},\ and\ \citenamefont {Zhu}}]{li2019metallic}%
  \BibitemOpen
  \bibfield  {author} {\bibinfo {author} {\bibfnamefont {Jing}\ \bibnamefont
  {Li}}, \bibinfo {author} {\bibfnamefont {Hailong}\ \bibnamefont {Fu}},
  \bibinfo {author} {\bibfnamefont {Zhenxi}\ \bibnamefont {Yin}}, \bibinfo
  {author} {\bibfnamefont {Kenji}\ \bibnamefont {Watanabe}}, \bibinfo {author}
  {\bibfnamefont {Takashi}\ \bibnamefont {Taniguchi}}, \ and\ \bibinfo {author}
  {\bibfnamefont {Jun}\ \bibnamefont {Zhu}},\ }\bibfield  {title} {\enquote
  {\bibinfo {title} {Metallic phase and temperature dependence of the $\nu$= 0
  quantum hall state in bilayer graphene},}\ }\href@noop {} {\bibfield
  {journal} {\bibinfo  {journal} {Physical review letters}\ }\textbf {\bibinfo
  {volume} {122}},\ \bibinfo {pages} {097701} (\bibinfo {year}
  {2019})}\BibitemShut {NoStop}%
\bibitem [{\citenamefont {Ohta}\ \emph {et~al.}(2006)\citenamefont {Ohta},
  \citenamefont {Bostwick}, \citenamefont {Seyller}, \citenamefont {Horn},\
  and\ \citenamefont {Rotenberg}}]{ohta2006controlling}%
  \BibitemOpen
  \bibfield  {author} {\bibinfo {author} {\bibfnamefont {Taisuke}\ \bibnamefont
  {Ohta}}, \bibinfo {author} {\bibfnamefont {Aaron}\ \bibnamefont {Bostwick}},
  \bibinfo {author} {\bibfnamefont {Thomas}\ \bibnamefont {Seyller}}, \bibinfo
  {author} {\bibfnamefont {Karsten}\ \bibnamefont {Horn}}, \ and\ \bibinfo
  {author} {\bibfnamefont {Eli}\ \bibnamefont {Rotenberg}},\ }\bibfield
  {title} {\enquote {\bibinfo {title} {Controlling the electronic structure of
  bilayer graphene},}\ }\href@noop {} {\bibfield  {journal} {\bibinfo
  {journal} {Science}\ }\textbf {\bibinfo {volume} {313}},\ \bibinfo {pages}
  {951--954} (\bibinfo {year} {2006})}\BibitemShut {NoStop}%
\bibitem [{\citenamefont {Park}\ \emph {et~al.}(2012)\citenamefont {Park},
  \citenamefont {Jo}, \citenamefont {Yu}, \citenamefont {Kim}, \citenamefont
  {Yang}, \citenamefont {Lee}, \citenamefont {Kim}, \citenamefont {Hong},
  \citenamefont {Kim}, \citenamefont {Cho} \emph {et~al.}}]{park2012single}%
  \BibitemOpen
  \bibfield  {author} {\bibinfo {author} {\bibfnamefont {Jaesung}\ \bibnamefont
  {Park}}, \bibinfo {author} {\bibfnamefont {Sae~Byeok}\ \bibnamefont {Jo}},
  \bibinfo {author} {\bibfnamefont {Young-Jun}\ \bibnamefont {Yu}}, \bibinfo
  {author} {\bibfnamefont {Youngsoo}\ \bibnamefont {Kim}}, \bibinfo {author}
  {\bibfnamefont {Jae~Won}\ \bibnamefont {Yang}}, \bibinfo {author}
  {\bibfnamefont {Wi~Hyoung}\ \bibnamefont {Lee}}, \bibinfo {author}
  {\bibfnamefont {Hyun~Ho}\ \bibnamefont {Kim}}, \bibinfo {author}
  {\bibfnamefont {Byung~Hee}\ \bibnamefont {Hong}}, \bibinfo {author}
  {\bibfnamefont {Philip}\ \bibnamefont {Kim}}, \bibinfo {author}
  {\bibfnamefont {Kilwon}\ \bibnamefont {Cho}},  \emph {et~al.},\ }\bibfield
  {title} {\enquote {\bibinfo {title} {Single-gate bandgap opening of bilayer
  graphene by dual molecular doping},}\ }\href@noop {} {\bibfield  {journal}
  {\bibinfo  {journal} {Advanced materials}\ }\textbf {\bibinfo {volume}
  {24}},\ \bibinfo {pages} {407--411} (\bibinfo {year} {2012})}\BibitemShut
  {NoStop}%
\bibitem [{\citenamefont {Mele}(2010)}]{mele2010commensuration}%
  \BibitemOpen
  \bibfield  {author} {\bibinfo {author} {\bibfnamefont {E.J.}\ \bibnamefont
  {Mele}},\ }\bibfield  {title} {\enquote {\bibinfo {title} {Commensuration and
  interlayer coherence in twisted bilayer graphene},}\ }\href@noop {}
  {\bibfield  {journal} {\bibinfo  {journal} {Physical Review B}\ }\textbf
  {\bibinfo {volume} {81}},\ \bibinfo {pages} {161405(R)} (\bibinfo {year}
  {2010})}\BibitemShut {NoStop}%
\bibitem [{\citenamefont {Shallcross}\ \emph {et~al.}(2010)\citenamefont
  {Shallcross}, \citenamefont {Sharma}, \citenamefont {Kandelaki},\ and\
  \citenamefont {Pankratov}}]{shallcross2010electronic}%
  \BibitemOpen
  \bibfield  {author} {\bibinfo {author} {\bibfnamefont {S}~\bibnamefont
  {Shallcross}}, \bibinfo {author} {\bibfnamefont {S}~\bibnamefont {Sharma}},
  \bibinfo {author} {\bibfnamefont {E}~\bibnamefont {Kandelaki}}, \ and\
  \bibinfo {author} {\bibfnamefont {O.A.}\ \bibnamefont {Pankratov}},\
  }\bibfield  {title} {\enquote {\bibinfo {title} {Electronic structure of
  turbostratic graphene},}\ }\href@noop {} {\bibfield  {journal} {\bibinfo
  {journal} {Physical Review B}\ }\textbf {\bibinfo {volume} {81}},\ \bibinfo
  {pages} {165105} (\bibinfo {year} {2010})}\BibitemShut {NoStop}%
\bibitem [{\citenamefont {Cao}\ \emph {et~al.}(2018{\natexlab{b}})\citenamefont
  {Cao}, \citenamefont {Fatemi}, \citenamefont {Demir}, \citenamefont {Fang},
  \citenamefont {Tomarken}, \citenamefont {Luo}, \citenamefont
  {Sanchez-Yamagishi}, \citenamefont {Watanabe}, \citenamefont {Taniguchi},
  \citenamefont {Kaxiras} \emph {et~al.}}]{cao2018correlated}%
  \BibitemOpen
  \bibfield  {author} {\bibinfo {author} {\bibfnamefont {Yuan}\ \bibnamefont
  {Cao}}, \bibinfo {author} {\bibfnamefont {Valla}\ \bibnamefont {Fatemi}},
  \bibinfo {author} {\bibfnamefont {Ahmet}\ \bibnamefont {Demir}}, \bibinfo
  {author} {\bibfnamefont {Shiang}\ \bibnamefont {Fang}}, \bibinfo {author}
  {\bibfnamefont {Spencer~L}\ \bibnamefont {Tomarken}}, \bibinfo {author}
  {\bibfnamefont {Jason~Y}\ \bibnamefont {Luo}}, \bibinfo {author}
  {\bibfnamefont {Javier~D}\ \bibnamefont {Sanchez-Yamagishi}}, \bibinfo
  {author} {\bibfnamefont {Kenji}\ \bibnamefont {Watanabe}}, \bibinfo {author}
  {\bibfnamefont {Takashi}\ \bibnamefont {Taniguchi}}, \bibinfo {author}
  {\bibfnamefont {Efthimios}\ \bibnamefont {Kaxiras}},  \emph {et~al.},\
  }\bibfield  {title} {\enquote {\bibinfo {title} {Correlated insulator
  behaviour at half-filling in magic-angle graphene superlattices},}\
  }\href@noop {} {\bibfield  {journal} {\bibinfo  {journal} {Nature}\ }\textbf
  {\bibinfo {volume} {556}},\ \bibinfo {pages} {80--84} (\bibinfo {year}
  {2018}{\natexlab{b}})}\BibitemShut {NoStop}%
\bibitem [{\citenamefont {Zhang}\ \emph {et~al.}(2009)\citenamefont {Zhang},
  \citenamefont {Tang}, \citenamefont {Girit}, \citenamefont {Hao},
  \citenamefont {Martin}, \citenamefont {Zettl}, \citenamefont {Crommie},
  \citenamefont {Shen},\ and\ \citenamefont {Wang}}]{zhang2009direct}%
  \BibitemOpen
  \bibfield  {author} {\bibinfo {author} {\bibfnamefont {Yuanbo}\ \bibnamefont
  {Zhang}}, \bibinfo {author} {\bibfnamefont {Tsung-Ta}\ \bibnamefont {Tang}},
  \bibinfo {author} {\bibfnamefont {Caglar}\ \bibnamefont {Girit}}, \bibinfo
  {author} {\bibfnamefont {Zhao}\ \bibnamefont {Hao}}, \bibinfo {author}
  {\bibfnamefont {Michael~C}\ \bibnamefont {Martin}}, \bibinfo {author}
  {\bibfnamefont {Alex}\ \bibnamefont {Zettl}}, \bibinfo {author}
  {\bibfnamefont {Michael~F}\ \bibnamefont {Crommie}}, \bibinfo {author}
  {\bibfnamefont {Y~Ron}\ \bibnamefont {Shen}}, \ and\ \bibinfo {author}
  {\bibfnamefont {Feng}\ \bibnamefont {Wang}},\ }\bibfield  {title} {\enquote
  {\bibinfo {title} {Direct observation of a widely tunable bandgap in bilayer
  graphene},}\ }\href@noop {} {\bibfield  {journal} {\bibinfo  {journal}
  {Nature}\ }\textbf {\bibinfo {volume} {459}},\ \bibinfo {pages} {820--823}
  (\bibinfo {year} {2009})}\BibitemShut {NoStop}%
\bibitem [{\citenamefont {Mak}\ \emph {et~al.}(2009)\citenamefont {Mak},
  \citenamefont {Lui}, \citenamefont {Shan},\ and\ \citenamefont
  {Heinz}}]{mak2009observation}%
  \BibitemOpen
  \bibfield  {author} {\bibinfo {author} {\bibfnamefont {Kin~Fai}\ \bibnamefont
  {Mak}}, \bibinfo {author} {\bibfnamefont {Chun~Hung}\ \bibnamefont {Lui}},
  \bibinfo {author} {\bibfnamefont {Jie}\ \bibnamefont {Shan}}, \ and\ \bibinfo
  {author} {\bibfnamefont {Tony~F}\ \bibnamefont {Heinz}},\ }\bibfield  {title}
  {\enquote {\bibinfo {title} {Observation of an electric-field-induced band
  gap in bilayer graphene by infrared spectroscopy},}\ }\href@noop {}
  {\bibfield  {journal} {\bibinfo  {journal} {Physical review letters}\
  }\textbf {\bibinfo {volume} {102}},\ \bibinfo {pages} {256405} (\bibinfo
  {year} {2009})}\BibitemShut {NoStop}%
\bibitem [{\citenamefont {McCann}\ \emph {et~al.}(2007)\citenamefont {McCann},
  \citenamefont {Abergel},\ and\ \citenamefont {Fal'ko}}]{mccann2007low}%
  \BibitemOpen
  \bibfield  {author} {\bibinfo {author} {\bibfnamefont {Edward}\ \bibnamefont
  {McCann}}, \bibinfo {author} {\bibfnamefont {David~SL}\ \bibnamefont
  {Abergel}}, \ and\ \bibinfo {author} {\bibfnamefont {Vladimir~I}\
  \bibnamefont {Fal'ko}},\ }\bibfield  {title} {\enquote {\bibinfo {title} {The
  low energy electronic band structure of bilayer graphene},}\ }\href@noop {}
  {\bibfield  {journal} {\bibinfo  {journal} {The European Physical Journal
  Special Topics}\ }\textbf {\bibinfo {volume} {148}},\ \bibinfo {pages}
  {91--103} (\bibinfo {year} {2007})}\BibitemShut {NoStop}%
\bibitem [{\citenamefont {McCann}(2006)}]{PhysRevB.74.161403}%
  \BibitemOpen
  \bibfield  {author} {\bibinfo {author} {\bibfnamefont {Edward}\ \bibnamefont
  {McCann}},\ }\bibfield  {title} {\enquote {\bibinfo {title} {Asymmetry gap in
  the electronic band structure of bilayer graphene},}\ }\href {\doibase
  10.1103/PhysRevB.74.161403} {\bibfield  {journal} {\bibinfo  {journal} {Phys.
  Rev. B}\ }\textbf {\bibinfo {volume} {74}},\ \bibinfo {pages} {161403(R)}
  (\bibinfo {year} {2006})}\BibitemShut {NoStop}%
\bibitem [{\citenamefont {Castro}\ \emph {et~al.}(2007)\citenamefont {Castro},
  \citenamefont {Novoselov}, \citenamefont {Morozov}, \citenamefont {Peres},
  \citenamefont {dos Santos}, \citenamefont {Nilsson}, \citenamefont {Guinea},
  \citenamefont {Geim},\ and\ \citenamefont {Neto}}]{PhysRevLett.99.216802}%
  \BibitemOpen
  \bibfield  {author} {\bibinfo {author} {\bibfnamefont {E.V.}\ \bibnamefont
  {Castro}}, \bibinfo {author} {\bibfnamefont {K.~S.}\ \bibnamefont
  {Novoselov}}, \bibinfo {author} {\bibfnamefont {S.~V.}\ \bibnamefont
  {Morozov}}, \bibinfo {author} {\bibfnamefont {N.~M.~R.}\ \bibnamefont
  {Peres}}, \bibinfo {author} {\bibfnamefont {J.~M. B.~Lopes}\ \bibnamefont
  {dos Santos}}, \bibinfo {author} {\bibfnamefont {Johan}\ \bibnamefont
  {Nilsson}}, \bibinfo {author} {\bibfnamefont {F.}~\bibnamefont {Guinea}},
  \bibinfo {author} {\bibfnamefont {A.~K.}\ \bibnamefont {Geim}}, \ and\
  \bibinfo {author} {\bibfnamefont {A.~H.~Castro}\ \bibnamefont {Neto}},\
  }\bibfield  {title} {\enquote {\bibinfo {title} {Biased bilayer graphene:
  Semiconductor with a gap tunable by the electric field effect},}\ }\href
  {\doibase 10.1103/PhysRevLett.99.216802} {\bibfield  {journal} {\bibinfo
  {journal} {Phys. Rev. Lett.}\ }\textbf {\bibinfo {volume} {99}},\ \bibinfo
  {pages} {216802} (\bibinfo {year} {2007})}\BibitemShut {NoStop}%
\bibitem [{\citenamefont {McCann}\ and\ \citenamefont
  {Koshino}(2013)}]{mccann2013electronic}%
  \BibitemOpen
  \bibfield  {author} {\bibinfo {author} {\bibfnamefont {Edward}\ \bibnamefont
  {McCann}}\ and\ \bibinfo {author} {\bibfnamefont {Mikito}\ \bibnamefont
  {Koshino}},\ }\bibfield  {title} {\enquote {\bibinfo {title} {The electronic
  properties of bilayer graphene},}\ }\href@noop {} {\bibfield  {journal}
  {\bibinfo  {journal} {Reports on Progress in physics}\ }\textbf {\bibinfo
  {volume} {76}},\ \bibinfo {pages} {056503} (\bibinfo {year}
  {2013})}\BibitemShut {NoStop}%
\bibitem [{\citenamefont {Weitz}\ \emph {et~al.}(2010)\citenamefont {Weitz},
  \citenamefont {Allen}, \citenamefont {Feldman}, \citenamefont {Martin},\ and\
  \citenamefont {Yacoby}}]{weitz2010broken}%
  \BibitemOpen
  \bibfield  {author} {\bibinfo {author} {\bibfnamefont {R~Thomas}\
  \bibnamefont {Weitz}}, \bibinfo {author} {\bibfnamefont {MT}~\bibnamefont
  {Allen}}, \bibinfo {author} {\bibfnamefont {BE}~\bibnamefont {Feldman}},
  \bibinfo {author} {\bibfnamefont {J}~\bibnamefont {Martin}}, \ and\ \bibinfo
  {author} {\bibfnamefont {A}~\bibnamefont {Yacoby}},\ }\bibfield  {title}
  {\enquote {\bibinfo {title} {Broken-symmetry states in doubly gated suspended
  bilayer graphene},}\ }\href@noop {} {\bibfield  {journal} {\bibinfo
  {journal} {Science}\ }\textbf {\bibinfo {volume} {330}},\ \bibinfo {pages}
  {812--816} (\bibinfo {year} {2010})}\BibitemShut {NoStop}%
\bibitem [{\citenamefont {Bao}\ \emph {et~al.}(2012)\citenamefont {Bao},
  \citenamefont {Velasco}, \citenamefont {Zhang}, \citenamefont {Jing},
  \citenamefont {Standley}, \citenamefont {Smirnov}, \citenamefont {Bockrath},
  \citenamefont {MacDonald},\ and\ \citenamefont {Lau}}]{bao2012evidence}%
  \BibitemOpen
  \bibfield  {author} {\bibinfo {author} {\bibfnamefont {Wenzhong}\
  \bibnamefont {Bao}}, \bibinfo {author} {\bibfnamefont {Jairo}\ \bibnamefont
  {Velasco}}, \bibinfo {author} {\bibfnamefont {Fan}\ \bibnamefont {Zhang}},
  \bibinfo {author} {\bibfnamefont {Lei}\ \bibnamefont {Jing}}, \bibinfo
  {author} {\bibfnamefont {Brian}\ \bibnamefont {Standley}}, \bibinfo {author}
  {\bibfnamefont {Dmitry}\ \bibnamefont {Smirnov}}, \bibinfo {author}
  {\bibfnamefont {Marc}\ \bibnamefont {Bockrath}}, \bibinfo {author}
  {\bibfnamefont {Allan~H}\ \bibnamefont {MacDonald}}, \ and\ \bibinfo {author}
  {\bibfnamefont {Chun~Ning}\ \bibnamefont {Lau}},\ }\bibfield  {title}
  {\enquote {\bibinfo {title} {Evidence for a spontaneous gapped state in
  ultraclean bilayer graphene},}\ }\href@noop {} {\bibfield  {journal}
  {\bibinfo  {journal} {Proceedings of the National Academy of Sciences}\
  }\textbf {\bibinfo {volume} {109}},\ \bibinfo {pages} {10802--10805}
  (\bibinfo {year} {2012})}\BibitemShut {NoStop}%
\bibitem [{\citenamefont {Velasco}\ \emph {et~al.}(2012)\citenamefont
  {Velasco}, \citenamefont {Jing}, \citenamefont {Bao}, \citenamefont {Lee},
  \citenamefont {Kratz}, \citenamefont {Aji}, \citenamefont {Bockrath},
  \citenamefont {Lau}, \citenamefont {Varma}, \citenamefont {Stillwell} \emph
  {et~al.}}]{velasco2012transport}%
  \BibitemOpen
  \bibfield  {author} {\bibinfo {author} {\bibfnamefont {Jr}~\bibnamefont
  {Velasco}}, \bibinfo {author} {\bibfnamefont {Lei}\ \bibnamefont {Jing}},
  \bibinfo {author} {\bibfnamefont {Wenzhong}\ \bibnamefont {Bao}}, \bibinfo
  {author} {\bibfnamefont {Yongjin}\ \bibnamefont {Lee}}, \bibinfo {author}
  {\bibfnamefont {Philip}\ \bibnamefont {Kratz}}, \bibinfo {author}
  {\bibfnamefont {Vivek}\ \bibnamefont {Aji}}, \bibinfo {author} {\bibfnamefont
  {Marc}\ \bibnamefont {Bockrath}}, \bibinfo {author} {\bibfnamefont
  {CN}~\bibnamefont {Lau}}, \bibinfo {author} {\bibfnamefont {Chandra}\
  \bibnamefont {Varma}}, \bibinfo {author} {\bibfnamefont {Ryan}\ \bibnamefont
  {Stillwell}},  \emph {et~al.},\ }\bibfield  {title} {\enquote {\bibinfo
  {title} {Transport spectroscopy of symmetry-broken insulating states in
  bilayer graphene},}\ }\href@noop {} {\bibfield  {journal} {\bibinfo
  {journal} {Nature nanotechnology}\ }\textbf {\bibinfo {volume} {7}},\
  \bibinfo {pages} {156--160} (\bibinfo {year} {2012})}\BibitemShut {NoStop}%
\bibitem [{\citenamefont {Lang}\ \emph {et~al.}(2012)\citenamefont {Lang},
  \citenamefont {Meng}, \citenamefont {Scherer}, \citenamefont {Uebelacker},
  \citenamefont {Assaad}, \citenamefont {Muramatsu}, \citenamefont
  {Honerkamp},\ and\ \citenamefont {Wessel}}]{lang2012antiferromagnetism}%
  \BibitemOpen
  \bibfield  {author} {\bibinfo {author} {\bibfnamefont {Thomas~C}\
  \bibnamefont {Lang}}, \bibinfo {author} {\bibfnamefont {Zi~Yang}\
  \bibnamefont {Meng}}, \bibinfo {author} {\bibfnamefont {Michael~M}\
  \bibnamefont {Scherer}}, \bibinfo {author} {\bibfnamefont {Stefan}\
  \bibnamefont {Uebelacker}}, \bibinfo {author} {\bibfnamefont {Fakher~F}\
  \bibnamefont {Assaad}}, \bibinfo {author} {\bibfnamefont {Alejandro}\
  \bibnamefont {Muramatsu}}, \bibinfo {author} {\bibfnamefont {Carsten}\
  \bibnamefont {Honerkamp}}, \ and\ \bibinfo {author} {\bibfnamefont {Stefan}\
  \bibnamefont {Wessel}},\ }\bibfield  {title} {\enquote {\bibinfo {title}
  {Antiferromagnetism in the hubbard model on the bernal-stacked honeycomb
  bilayer},}\ }\href@noop {} {\bibfield  {journal} {\bibinfo  {journal}
  {Physical review letters}\ }\textbf {\bibinfo {volume} {109}},\ \bibinfo
  {pages} {126402} (\bibinfo {year} {2012})}\BibitemShut {NoStop}%
\bibitem [{\citenamefont {Yuan}\ \emph {et~al.}(2013)\citenamefont {Yuan},
  \citenamefont {Xu}, \citenamefont {Wang}, \citenamefont {Zhou}, \citenamefont
  {Gao},\ and\ \citenamefont {Zhang}}]{yuan2013possible}%
  \BibitemOpen
  \bibfield  {author} {\bibinfo {author} {\bibfnamefont {Jie}\ \bibnamefont
  {Yuan}}, \bibinfo {author} {\bibfnamefont {Dong-Hui}\ \bibnamefont {Xu}},
  \bibinfo {author} {\bibfnamefont {Hao}\ \bibnamefont {Wang}}, \bibinfo
  {author} {\bibfnamefont {Yi}~\bibnamefont {Zhou}}, \bibinfo {author}
  {\bibfnamefont {Jin-Hua}\ \bibnamefont {Gao}}, \ and\ \bibinfo {author}
  {\bibfnamefont {Fu-Chun}\ \bibnamefont {Zhang}},\ }\bibfield  {title}
  {\enquote {\bibinfo {title} {Possible half-metallic phase in bilayer
  graphene: Calculations based on mean-field theory applied to a two-layer
  hubbard model},}\ }\href@noop {} {\bibfield  {journal} {\bibinfo  {journal}
  {Physical Review B}\ }\textbf {\bibinfo {volume} {88}},\ \bibinfo {pages}
  {201109(R)} (\bibinfo {year} {2013})}\BibitemShut {NoStop}%
\bibitem [{\citenamefont {Jung}\ \emph {et~al.}(2011)\citenamefont {Jung},
  \citenamefont {Zhang},\ and\ \citenamefont {MacDonald}}]{jung2011lattice}%
  \BibitemOpen
  \bibfield  {author} {\bibinfo {author} {\bibfnamefont {Jeil}\ \bibnamefont
  {Jung}}, \bibinfo {author} {\bibfnamefont {Fan}\ \bibnamefont {Zhang}}, \
  and\ \bibinfo {author} {\bibfnamefont {Allan~H}\ \bibnamefont {MacDonald}},\
  }\bibfield  {title} {\enquote {\bibinfo {title} {Lattice theory of pseudospin
  ferromagnetism in bilayer graphene: Competing interaction-induced quantum
  hall states},}\ }\href@noop {} {\bibfield  {journal} {\bibinfo  {journal}
  {Physical Review B}\ }\textbf {\bibinfo {volume} {83}},\ \bibinfo {pages}
  {115408} (\bibinfo {year} {2011})}\BibitemShut {NoStop}%
\bibitem [{\citenamefont {Wang}\ \emph {et~al.}(2013)\citenamefont {Wang},
  \citenamefont {Wang}, \citenamefont {Gao},\ and\ \citenamefont
  {Zhang}}]{wang2013layer}%
  \BibitemOpen
  \bibfield  {author} {\bibinfo {author} {\bibfnamefont {Yong}\ \bibnamefont
  {Wang}}, \bibinfo {author} {\bibfnamefont {Hao}\ \bibnamefont {Wang}},
  \bibinfo {author} {\bibfnamefont {Jin-Hua}\ \bibnamefont {Gao}}, \ and\
  \bibinfo {author} {\bibfnamefont {Fu-Chun}\ \bibnamefont {Zhang}},\
  }\bibfield  {title} {\enquote {\bibinfo {title} {Layer antiferromagnetic
  state in bilayer graphene: A first-principles investigation},}\ }\href@noop
  {} {\bibfield  {journal} {\bibinfo  {journal} {Physical Review B}\ }\textbf
  {\bibinfo {volume} {87}},\ \bibinfo {pages} {195413} (\bibinfo {year}
  {2013})}\BibitemShut {NoStop}%
\bibitem [{\citenamefont {Scherer}\ \emph {et~al.}(2012)\citenamefont
  {Scherer}, \citenamefont {Uebelacker},\ and\ \citenamefont
  {Honerkamp}}]{scherer2012instabilities}%
  \BibitemOpen
  \bibfield  {author} {\bibinfo {author} {\bibfnamefont {Michael~M}\
  \bibnamefont {Scherer}}, \bibinfo {author} {\bibfnamefont {Stefan}\
  \bibnamefont {Uebelacker}}, \ and\ \bibinfo {author} {\bibfnamefont
  {Carsten}\ \bibnamefont {Honerkamp}},\ }\bibfield  {title} {\enquote
  {\bibinfo {title} {Instabilities of interacting electrons on the honeycomb
  bilayer},}\ }\href@noop {} {\bibfield  {journal} {\bibinfo  {journal}
  {Physical Review B}\ }\textbf {\bibinfo {volume} {85}},\ \bibinfo {pages}
  {235408} (\bibinfo {year} {2012})}\BibitemShut {NoStop}%
\bibitem [{\citenamefont {Kharitonov}(2012)}]{kharitonov2012canted}%
  \BibitemOpen
  \bibfield  {author} {\bibinfo {author} {\bibfnamefont {Maxim}\ \bibnamefont
  {Kharitonov}},\ }\bibfield  {title} {\enquote {\bibinfo {title} {Canted
  antiferromagnetic phase of the $\nu$= 0 quantum hall state in bilayer
  graphene},}\ }\href@noop {} {\bibfield  {journal} {\bibinfo  {journal}
  {Physical review letters}\ }\textbf {\bibinfo {volume} {109}},\ \bibinfo
  {pages} {046803} (\bibinfo {year} {2012})}\BibitemShut {NoStop}%
\bibitem [{\citenamefont {Lemonik}\ \emph {et~al.}(2012)\citenamefont
  {Lemonik}, \citenamefont {Aleiner},\ and\ \citenamefont
  {Fal'Ko}}]{lemonik2012competing}%
  \BibitemOpen
  \bibfield  {author} {\bibinfo {author} {\bibfnamefont {Y}~\bibnamefont
  {Lemonik}}, \bibinfo {author} {\bibfnamefont {I}~\bibnamefont {Aleiner}}, \
  and\ \bibinfo {author} {\bibfnamefont {VI}~\bibnamefont {Fal'Ko}},\
  }\bibfield  {title} {\enquote {\bibinfo {title} {Competing nematic,
  antiferromagnetic, and spin-flux orders in the ground state of bilayer
  graphene},}\ }\href@noop {} {\bibfield  {journal} {\bibinfo  {journal}
  {Physical review b}\ }\textbf {\bibinfo {volume} {85}},\ \bibinfo {pages}
  {245451} (\bibinfo {year} {2012})}\BibitemShut {NoStop}%
\bibitem [{\citenamefont {Nandkishore}\ and\ \citenamefont
  {Levitov}(2010)}]{nandkishore2010quantum}%
  \BibitemOpen
  \bibfield  {author} {\bibinfo {author} {\bibfnamefont {Rahul}\ \bibnamefont
  {Nandkishore}}\ and\ \bibinfo {author} {\bibfnamefont {Leonid}\ \bibnamefont
  {Levitov}},\ }\bibfield  {title} {\enquote {\bibinfo {title} {Quantum
  anomalous hall state in bilayer graphene},}\ }\href@noop {} {\bibfield
  {journal} {\bibinfo  {journal} {Physical Review B}\ }\textbf {\bibinfo
  {volume} {82}},\ \bibinfo {pages} {115124} (\bibinfo {year}
  {2010})}\BibitemShut {NoStop}%
\bibitem [{\citenamefont {Ray}\ and\ \citenamefont
  {Janssen}(2021)}]{ray2021gross}%
  \BibitemOpen
  \bibfield  {author} {\bibinfo {author} {\bibfnamefont {Shouryya}\
  \bibnamefont {Ray}}\ and\ \bibinfo {author} {\bibfnamefont {Lukas}\
  \bibnamefont {Janssen}},\ }\bibfield  {title} {\enquote {\bibinfo {title}
  {Gross-neveu-heisenberg criticality from competing nematic and
  antiferromagnetic orders in bilayer graphene},}\ }\href@noop {} {\bibfield
  {journal} {\bibinfo  {journal} {Physical Review B}\ }\textbf {\bibinfo
  {volume} {104}},\ \bibinfo {pages} {045101} (\bibinfo {year}
  {2021})}\BibitemShut {NoStop}%
\bibitem [{\citenamefont {Xu}\ \emph {et~al.}(2016)\citenamefont {Xu},
  \citenamefont {Song}, \citenamefont {Lin},\ and\ \citenamefont
  {Zhang}}]{xu2016gate}%
  \BibitemOpen
  \bibfield  {author} {\bibinfo {author} {\bibfnamefont {Jin-Rong}\
  \bibnamefont {Xu}}, \bibinfo {author} {\bibfnamefont {Ze-Yi}\ \bibnamefont
  {Song}}, \bibinfo {author} {\bibfnamefont {Hai-Qing}\ \bibnamefont {Lin}}, \
  and\ \bibinfo {author} {\bibfnamefont {Yu-Zhong}\ \bibnamefont {Zhang}},\
  }\bibfield  {title} {\enquote {\bibinfo {title} {Gate-induced gap in bilayer
  graphene suppressed by coulomb repulsion},}\ }\href@noop {} {\bibfield
  {journal} {\bibinfo  {journal} {Physical Review B}\ }\textbf {\bibinfo
  {volume} {93}},\ \bibinfo {pages} {035109} (\bibinfo {year}
  {2016})}\BibitemShut {NoStop}%
\bibitem [{\citenamefont {Cvetkovic}\ \emph {et~al.}(2012)\citenamefont
  {Cvetkovic}, \citenamefont {Throckmorton},\ and\ \citenamefont
  {Vafek}}]{PhysRevB.86.075467}%
  \BibitemOpen
  \bibfield  {author} {\bibinfo {author} {\bibfnamefont {Vladimir}\
  \bibnamefont {Cvetkovic}}, \bibinfo {author} {\bibfnamefont {Robert~E.}\
  \bibnamefont {Throckmorton}}, \ and\ \bibinfo {author} {\bibfnamefont
  {Oskar}\ \bibnamefont {Vafek}},\ }\bibfield  {title} {\enquote {\bibinfo
  {title} {Electronic multicriticality in bilayer graphene},}\ }\href {\doibase
  10.1103/PhysRevB.86.075467} {\bibfield  {journal} {\bibinfo  {journal} {Phys.
  Rev. B}\ }\textbf {\bibinfo {volume} {86}},\ \bibinfo {pages} {075467}
  (\bibinfo {year} {2012})}\BibitemShut {NoStop}%
\bibitem [{\citenamefont {Throckmorton}\ and\ \citenamefont
  {Vafek}(2012)}]{PhysRevB.86.115447}%
  \BibitemOpen
  \bibfield  {author} {\bibinfo {author} {\bibfnamefont {Robert~E.}\
  \bibnamefont {Throckmorton}}\ and\ \bibinfo {author} {\bibfnamefont {Oskar}\
  \bibnamefont {Vafek}},\ }\bibfield  {title} {\enquote {\bibinfo {title}
  {Fermions on bilayer graphene: Symmetry breaking for $b=0$ and
  $\ensuremath{\nu}=0$},}\ }\href {\doibase 10.1103/PhysRevB.86.115447}
  {\bibfield  {journal} {\bibinfo  {journal} {Phys. Rev. B}\ }\textbf {\bibinfo
  {volume} {86}},\ \bibinfo {pages} {115447} (\bibinfo {year}
  {2012})}\BibitemShut {NoStop}%
\bibitem [{\citenamefont {Szab\'o}\ and\ \citenamefont
  {Roy}(2021)}]{PhysRevB.103.205135}%
  \BibitemOpen
  \bibfield  {author} {\bibinfo {author} {\bibfnamefont {Andr\'as~L.}\
  \bibnamefont {Szab\'o}}\ and\ \bibinfo {author} {\bibfnamefont {Bitan}\
  \bibnamefont {Roy}},\ }\bibfield  {title} {\enquote {\bibinfo {title}
  {Extended hubbard model in undoped and doped monolayer and bilayer graphene:
  Selection rules and organizing principle among competing orders},}\ }\href
  {\doibase 10.1103/PhysRevB.103.205135} {\bibfield  {journal} {\bibinfo
  {journal} {Phys. Rev. B}\ }\textbf {\bibinfo {volume} {103}},\ \bibinfo
  {pages} {205135} (\bibinfo {year} {2021})}\BibitemShut {NoStop}%
\bibitem [{\citenamefont {Zhang}\ \emph {et~al.}(2011)\citenamefont {Zhang},
  \citenamefont {Jung}, \citenamefont {Fiete}, \citenamefont {Niu},\ and\
  \citenamefont {MacDonald}}]{zhang2011spontaneous}%
  \BibitemOpen
  \bibfield  {author} {\bibinfo {author} {\bibfnamefont {Fan}\ \bibnamefont
  {Zhang}}, \bibinfo {author} {\bibfnamefont {Jeil}\ \bibnamefont {Jung}},
  \bibinfo {author} {\bibfnamefont {Gregory~A}\ \bibnamefont {Fiete}}, \bibinfo
  {author} {\bibfnamefont {Qian}\ \bibnamefont {Niu}}, \ and\ \bibinfo {author}
  {\bibfnamefont {Allan~H}\ \bibnamefont {MacDonald}},\ }\bibfield  {title}
  {\enquote {\bibinfo {title} {Spontaneous quantum hall states in chirally
  stacked few-layer graphene systems},}\ }\href@noop {} {\bibfield  {journal}
  {\bibinfo  {journal} {Physical review letters}\ }\textbf {\bibinfo {volume}
  {106}},\ \bibinfo {pages} {156801} (\bibinfo {year} {2011})}\BibitemShut
  {NoStop}%
\bibitem [{\citenamefont {Stolyarova}\ \emph {et~al.}(2007)\citenamefont
  {Stolyarova}, \citenamefont {Rim}, \citenamefont {Ryu}, \citenamefont
  {Maultzsch}, \citenamefont {Kim}, \citenamefont {Brus}, \citenamefont
  {Heinz}, \citenamefont {Hybertsen},\ and\ \citenamefont
  {Flynn}}]{stolyarova2007high}%
  \BibitemOpen
  \bibfield  {author} {\bibinfo {author} {\bibfnamefont {Elena}\ \bibnamefont
  {Stolyarova}}, \bibinfo {author} {\bibfnamefont {Kwang~Taeg}\ \bibnamefont
  {Rim}}, \bibinfo {author} {\bibfnamefont {Sunmin}\ \bibnamefont {Ryu}},
  \bibinfo {author} {\bibfnamefont {Janina}\ \bibnamefont {Maultzsch}},
  \bibinfo {author} {\bibfnamefont {Philip}\ \bibnamefont {Kim}}, \bibinfo
  {author} {\bibfnamefont {Louis~E}\ \bibnamefont {Brus}}, \bibinfo {author}
  {\bibfnamefont {Tony~F}\ \bibnamefont {Heinz}}, \bibinfo {author}
  {\bibfnamefont {Mark~S}\ \bibnamefont {Hybertsen}}, \ and\ \bibinfo {author}
  {\bibfnamefont {George~W}\ \bibnamefont {Flynn}},\ }\bibfield  {title}
  {\enquote {\bibinfo {title} {High-resolution scanning tunneling microscopy
  imaging of mesoscopic graphene sheets on an insulating surface},}\
  }\href@noop {} {\bibfield  {journal} {\bibinfo  {journal} {Proceedings of the
  National Academy of Sciences}\ }\textbf {\bibinfo {volume} {104}},\ \bibinfo
  {pages} {9209--9212} (\bibinfo {year} {2007})}\BibitemShut {NoStop}%
\bibitem [{\citenamefont {Meyer}\ \emph {et~al.}(2007)\citenamefont {Meyer},
  \citenamefont {Geim}, \citenamefont {Katsnelson}, \citenamefont {Novoselov},
  \citenamefont {Booth},\ and\ \citenamefont {Roth}}]{meyer2007structure}%
  \BibitemOpen
  \bibfield  {author} {\bibinfo {author} {\bibfnamefont {Jannik~C}\
  \bibnamefont {Meyer}}, \bibinfo {author} {\bibfnamefont {Andre~K}\
  \bibnamefont {Geim}}, \bibinfo {author} {\bibfnamefont {Mikhail~I}\
  \bibnamefont {Katsnelson}}, \bibinfo {author} {\bibfnamefont {Konstantin~S}\
  \bibnamefont {Novoselov}}, \bibinfo {author} {\bibfnamefont {Tim~J}\
  \bibnamefont {Booth}}, \ and\ \bibinfo {author} {\bibfnamefont {Siegmar}\
  \bibnamefont {Roth}},\ }\bibfield  {title} {\enquote {\bibinfo {title} {The
  structure of suspended graphene sheets},}\ }\href@noop {} {\bibfield
  {journal} {\bibinfo  {journal} {Nature}\ }\textbf {\bibinfo {volume} {446}},\
  \bibinfo {pages} {60--63} (\bibinfo {year} {2007})}\BibitemShut {NoStop}%
\bibitem [{\citenamefont {Fasolino}\ \emph {et~al.}(2007)\citenamefont
  {Fasolino}, \citenamefont {Los},\ and\ \citenamefont
  {Katsnelson}}]{fasolino2007intrinsic}%
  \BibitemOpen
  \bibfield  {author} {\bibinfo {author} {\bibfnamefont {Annalisa}\
  \bibnamefont {Fasolino}}, \bibinfo {author} {\bibfnamefont {JH}~\bibnamefont
  {Los}}, \ and\ \bibinfo {author} {\bibfnamefont {Mikhail~I}\ \bibnamefont
  {Katsnelson}},\ }\bibfield  {title} {\enquote {\bibinfo {title} {Intrinsic
  ripples in graphene},}\ }\href@noop {} {\bibfield  {journal} {\bibinfo
  {journal} {Nature materials}\ }\textbf {\bibinfo {volume} {6}},\ \bibinfo
  {pages} {858--861} (\bibinfo {year} {2007})}\BibitemShut {NoStop}%
\bibitem [{\citenamefont {Zhang}\ \emph {et~al.}(2010)\citenamefont {Zhang},
  \citenamefont {Sahu}, \citenamefont {Min},\ and\ \citenamefont
  {MacDonald}}]{zhang2010band}%
  \BibitemOpen
  \bibfield  {author} {\bibinfo {author} {\bibfnamefont {Fan}\ \bibnamefont
  {Zhang}}, \bibinfo {author} {\bibfnamefont {Bhagawan}\ \bibnamefont {Sahu}},
  \bibinfo {author} {\bibfnamefont {Hongki}\ \bibnamefont {Min}}, \ and\
  \bibinfo {author} {\bibfnamefont {Allan~H}\ \bibnamefont {MacDonald}},\
  }\bibfield  {title} {\enquote {\bibinfo {title} {Band structure of
  abc-stacked graphene trilayers},}\ }\href@noop {} {\bibfield  {journal}
  {\bibinfo  {journal} {Physical Review B}\ }\textbf {\bibinfo {volume} {82}},\
  \bibinfo {pages} {035409} (\bibinfo {year} {2010})}\BibitemShut {NoStop}%
\bibitem [{\citenamefont {Gui}\ \emph {et~al.}(2012)\citenamefont {Gui},
  \citenamefont {Zhong},\ and\ \citenamefont {Ma}}]{gui2012electronic}%
  \BibitemOpen
  \bibfield  {author} {\bibinfo {author} {\bibfnamefont {Gui}\ \bibnamefont
  {Gui}}, \bibinfo {author} {\bibfnamefont {Jianxin}\ \bibnamefont {Zhong}}, \
  and\ \bibinfo {author} {\bibfnamefont {Zhenqiang}\ \bibnamefont {Ma}},\
  }\bibfield  {title} {\enquote {\bibinfo {title} {Electronic properties of
  rippled graphene},}\ }in\ \href@noop {} {\emph {\bibinfo {booktitle} {Journal
  of Physics: Conference Series}}},\ Vol.\ \bibinfo {volume} {402}\ (\bibinfo
  {organization} {IOP Publishing},\ \bibinfo {year} {2012})\ p.\ \bibinfo
  {pages} {012004}\BibitemShut {NoStop}%
\bibitem [{\citenamefont {Lin}\ \emph {et~al.}(2015)\citenamefont {Lin},
  \citenamefont {Chang}, \citenamefont {Shyu}, \citenamefont {Lu},\ and\
  \citenamefont {Lin}}]{lin2015feature}%
  \BibitemOpen
  \bibfield  {author} {\bibinfo {author} {\bibfnamefont {Shih-Yang}\
  \bibnamefont {Lin}}, \bibinfo {author} {\bibfnamefont {Shen-Lin}\
  \bibnamefont {Chang}}, \bibinfo {author} {\bibfnamefont {Feng-Lin}\
  \bibnamefont {Shyu}}, \bibinfo {author} {\bibfnamefont {Jian-Ming}\
  \bibnamefont {Lu}}, \ and\ \bibinfo {author} {\bibfnamefont {Ming-Fa}\
  \bibnamefont {Lin}},\ }\bibfield  {title} {\enquote {\bibinfo {title}
  {Feature-rich electronic properties in graphene ripples},}\ }\href@noop {}
  {\bibfield  {journal} {\bibinfo  {journal} {Carbon}\ }\textbf {\bibinfo
  {volume} {86}},\ \bibinfo {pages} {207--216} (\bibinfo {year}
  {2015})}\BibitemShut {NoStop}%
\bibitem [{\citenamefont {Jung}\ and\ \citenamefont
  {MacDonald}(2009)}]{jung2009theory}%
  \BibitemOpen
  \bibfield  {author} {\bibinfo {author} {\bibfnamefont {Jeil}\ \bibnamefont
  {Jung}}\ and\ \bibinfo {author} {\bibfnamefont {AH}~\bibnamefont
  {MacDonald}},\ }\bibfield  {title} {\enquote {\bibinfo {title} {Theory of the
  magnetic-field-induced insulator in neutral graphene sheets},}\ }\href@noop
  {} {\bibfield  {journal} {\bibinfo  {journal} {Physical Review B}\ }\textbf
  {\bibinfo {volume} {80}},\ \bibinfo {pages} {235417} (\bibinfo {year}
  {2009})}\BibitemShut {NoStop}%
\bibitem [{\citenamefont {Dion}\ \emph {et~al.}(2004)\citenamefont {Dion},
  \citenamefont {Rydberg}, \citenamefont {Schr{\"o}der}, \citenamefont
  {Langreth},\ and\ \citenamefont {Lundqvist}}]{dion2004van}%
  \BibitemOpen
  \bibfield  {author} {\bibinfo {author} {\bibfnamefont {Max}\ \bibnamefont
  {Dion}}, \bibinfo {author} {\bibfnamefont {Henrik}\ \bibnamefont {Rydberg}},
  \bibinfo {author} {\bibfnamefont {Elsebeth}\ \bibnamefont {Schr{\"o}der}},
  \bibinfo {author} {\bibfnamefont {David~C}\ \bibnamefont {Langreth}}, \ and\
  \bibinfo {author} {\bibfnamefont {Bengt~I}\ \bibnamefont {Lundqvist}},\
  }\bibfield  {title} {\enquote {\bibinfo {title} {Van der waals density
  functional for general geometries},}\ }\href@noop {} {\bibfield  {journal}
  {\bibinfo  {journal} {Physical review letters}\ }\textbf {\bibinfo {volume}
  {92}},\ \bibinfo {pages} {246401} (\bibinfo {year} {2004})}\BibitemShut
  {NoStop}%
\bibitem [{\citenamefont {Tkatchenko}\ and\ \citenamefont
  {Scheffler}(2009)}]{tkatchenko2009accurate}%
  \BibitemOpen
  \bibfield  {author} {\bibinfo {author} {\bibfnamefont {Alexandre}\
  \bibnamefont {Tkatchenko}}\ and\ \bibinfo {author} {\bibfnamefont {Matthias}\
  \bibnamefont {Scheffler}},\ }\bibfield  {title} {\enquote {\bibinfo {title}
  {Accurate molecular van der waals interactions from ground-state electron
  density and free-atom reference data},}\ }\href@noop {} {\bibfield  {journal}
  {\bibinfo  {journal} {Physical review letters}\ }\textbf {\bibinfo {volume}
  {102}},\ \bibinfo {pages} {073005} (\bibinfo {year} {2009})}\BibitemShut
  {NoStop}%
\bibitem [{\citenamefont {Cooper}(2010)}]{cooper2010van}%
  \BibitemOpen
  \bibfield  {author} {\bibinfo {author} {\bibfnamefont {Valentino~R}\
  \bibnamefont {Cooper}},\ }\bibfield  {title} {\enquote {\bibinfo {title} {Van
  der waals density functional: An appropriate exchange functional},}\
  }\href@noop {} {\bibfield  {journal} {\bibinfo  {journal} {Physical Review
  B}\ }\textbf {\bibinfo {volume} {81}},\ \bibinfo {pages} {161104} (\bibinfo
  {year} {2010})}\BibitemShut {NoStop}%
\bibitem [{\citenamefont {Grimme}\ \emph {et~al.}(2010)\citenamefont {Grimme},
  \citenamefont {Antony}, \citenamefont {Ehrlich},\ and\ \citenamefont
  {Krieg}}]{grimme2010consistent}%
  \BibitemOpen
  \bibfield  {author} {\bibinfo {author} {\bibfnamefont {Stefan}\ \bibnamefont
  {Grimme}}, \bibinfo {author} {\bibfnamefont {Jens}\ \bibnamefont {Antony}},
  \bibinfo {author} {\bibfnamefont {Stephan}\ \bibnamefont {Ehrlich}}, \ and\
  \bibinfo {author} {\bibfnamefont {Helge}\ \bibnamefont {Krieg}},\ }\bibfield
  {title} {\enquote {\bibinfo {title} {A consistent and accurate ab initio
  parametrization of density functional dispersion correction (dft-d) for the
  94 elements h-pu},}\ }\href@noop {} {\bibfield  {journal} {\bibinfo
  {journal} {The Journal of chemical physics}\ }\textbf {\bibinfo {volume}
  {132}},\ \bibinfo {pages} {154104} (\bibinfo {year} {2010})}\BibitemShut
  {NoStop}%
\bibitem [{\citenamefont {Marzari}\ \emph {et~al.}(2012)\citenamefont
  {Marzari}, \citenamefont {Mostofi}, \citenamefont {Yates}, \citenamefont
  {Souza},\ and\ \citenamefont {Vanderbilt}}]{marzari2012maximally}%
  \BibitemOpen
  \bibfield  {author} {\bibinfo {author} {\bibfnamefont {Nicola}\ \bibnamefont
  {Marzari}}, \bibinfo {author} {\bibfnamefont {Arash~A}\ \bibnamefont
  {Mostofi}}, \bibinfo {author} {\bibfnamefont {Jonathan~R}\ \bibnamefont
  {Yates}}, \bibinfo {author} {\bibfnamefont {Ivo}\ \bibnamefont {Souza}}, \
  and\ \bibinfo {author} {\bibfnamefont {David}\ \bibnamefont {Vanderbilt}},\
  }\bibfield  {title} {\enquote {\bibinfo {title} {Maximally localized wannier
  functions: Theory and applications},}\ }\href@noop {} {\bibfield  {journal}
  {\bibinfo  {journal} {Reviews of Modern Physics}\ }\textbf {\bibinfo {volume}
  {84}},\ \bibinfo {pages} {1419} (\bibinfo {year} {2012})}\BibitemShut
  {NoStop}%
\bibitem [{\citenamefont {Mostofi}\ \emph {et~al.}(2008)\citenamefont
  {Mostofi}, \citenamefont {Yates}, \citenamefont {Lee}, \citenamefont {Souza},
  \citenamefont {Vanderbilt},\ and\ \citenamefont
  {Marzari}}]{mostofi2008wannier90}%
  \BibitemOpen
  \bibfield  {author} {\bibinfo {author} {\bibfnamefont {Arash~A}\ \bibnamefont
  {Mostofi}}, \bibinfo {author} {\bibfnamefont {Jonathan~R}\ \bibnamefont
  {Yates}}, \bibinfo {author} {\bibfnamefont {Young-Su}\ \bibnamefont {Lee}},
  \bibinfo {author} {\bibfnamefont {Ivo}\ \bibnamefont {Souza}}, \bibinfo
  {author} {\bibfnamefont {David}\ \bibnamefont {Vanderbilt}}, \ and\ \bibinfo
  {author} {\bibfnamefont {Nicola}\ \bibnamefont {Marzari}},\ }\bibfield
  {title} {\enquote {\bibinfo {title} {wannier90: A tool for obtaining
  maximally-localised wannier functions},}\ }\href@noop {} {\bibfield
  {journal} {\bibinfo  {journal} {Computer physics communications}\ }\textbf
  {\bibinfo {volume} {178}},\ \bibinfo {pages} {685--699} (\bibinfo {year}
  {2008})}\BibitemShut {NoStop}%
\bibitem [{\citenamefont {Singh}\ and\ \citenamefont
  {Nordstrom}(2006)}]{singh2006planewaves}%
  \BibitemOpen
  \bibfield  {author} {\bibinfo {author} {\bibfnamefont {David~J}\ \bibnamefont
  {Singh}}\ and\ \bibinfo {author} {\bibfnamefont {Lars}\ \bibnamefont
  {Nordstrom}},\ }\href@noop {} {\emph {\bibinfo {title} {{Planewaves,
  Pseudopotentials, and the LAPW method}}}}\ (\bibinfo  {publisher} {Springer
  Science \& Business Media},\ \bibinfo {year} {2006})\BibitemShut {NoStop}%
\bibitem [{\citenamefont {Blaha}\ \emph {et~al.}(2001)\citenamefont {Blaha},
  \citenamefont {Schwarz}, \citenamefont {Madsen}, \citenamefont {Kvasnicka},
  \citenamefont {Luitz} \emph {et~al.}}]{blaha2001wien2k}%
  \BibitemOpen
  \bibfield  {author} {\bibinfo {author} {\bibfnamefont {Peter}\ \bibnamefont
  {Blaha}}, \bibinfo {author} {\bibfnamefont {Karlheinz}\ \bibnamefont
  {Schwarz}}, \bibinfo {author} {\bibfnamefont {Georg~KH}\ \bibnamefont
  {Madsen}}, \bibinfo {author} {\bibfnamefont {Dieter}\ \bibnamefont
  {Kvasnicka}}, \bibinfo {author} {\bibfnamefont {Joachim}\ \bibnamefont
  {Luitz}},  \emph {et~al.},\ }\bibfield  {title} {\enquote {\bibinfo {title}
  {{wien2k}},}\ }\href@noop {} {\bibfield  {journal} {\bibinfo  {journal} {An
  augmented plane wave+ local orbitals program for calculating crystal
  properties}\ }\textbf {\bibinfo {volume} {60}} (\bibinfo {year}
  {2001})}\BibitemShut {NoStop}%
\bibitem [{\citenamefont {Perdew}\ and\ \citenamefont
  {Zunger}(1981)}]{perdew1981self}%
  \BibitemOpen
  \bibfield  {author} {\bibinfo {author} {\bibfnamefont {John~P}\ \bibnamefont
  {Perdew}}\ and\ \bibinfo {author} {\bibfnamefont {Alex}\ \bibnamefont
  {Zunger}},\ }\bibfield  {title} {\enquote {\bibinfo {title}
  {{Self-interaction correction to density-functional approximations for
  many-electron systems}},}\ }\href@noop {} {\bibfield  {journal} {\bibinfo
  {journal} {Physical Review B}\ }\textbf {\bibinfo {volume} {23}},\ \bibinfo
  {pages} {5048} (\bibinfo {year} {1981})}\BibitemShut {NoStop}%
\bibitem [{\citenamefont {Bl{\"o}chl}\ \emph {et~al.}(1994)\citenamefont
  {Bl{\"o}chl}, \citenamefont {Jepsen},\ and\ \citenamefont
  {Andersen}}]{blochl1994improved}%
  \BibitemOpen
  \bibfield  {author} {\bibinfo {author} {\bibfnamefont {Peter~E}\ \bibnamefont
  {Bl{\"o}chl}}, \bibinfo {author} {\bibfnamefont {Ove}\ \bibnamefont
  {Jepsen}}, \ and\ \bibinfo {author} {\bibfnamefont {Ole~Krogh}\ \bibnamefont
  {Andersen}},\ }\bibfield  {title} {\enquote {\bibinfo {title} {{Improved
  tetrahedron method for Brillouin-zone integrations}},}\ }\href@noop {}
  {\bibfield  {journal} {\bibinfo  {journal} {Physical Review B}\ }\textbf
  {\bibinfo {volume} {49}},\ \bibinfo {pages} {16223} (\bibinfo {year}
  {1994})}\BibitemShut {NoStop}%
\bibitem [{\citenamefont {Freitag}\ \emph {et~al.}(2012)\citenamefont
  {Freitag}, \citenamefont {Trbovic}, \citenamefont {Weiss},\ and\
  \citenamefont {Sch{\"o}nenberger}}]{freitag2012spontaneously}%
  \BibitemOpen
  \bibfield  {author} {\bibinfo {author} {\bibfnamefont {F}~\bibnamefont
  {Freitag}}, \bibinfo {author} {\bibfnamefont {J}~\bibnamefont {Trbovic}},
  \bibinfo {author} {\bibfnamefont {M}~\bibnamefont {Weiss}}, \ and\ \bibinfo
  {author} {\bibfnamefont {C}~\bibnamefont {Sch{\"o}nenberger}},\ }\bibfield
  {title} {\enquote {\bibinfo {title} {Spontaneously gapped ground state in
  suspended bilayer graphene},}\ }\href@noop {} {\bibfield  {journal} {\bibinfo
   {journal} {Physical review letters}\ }\textbf {\bibinfo {volume} {108}},\
  \bibinfo {pages} {076602} (\bibinfo {year} {2012})}\BibitemShut {NoStop}%
\bibitem [{\citenamefont {Veligura}\ \emph {et~al.}(2012)\citenamefont
  {Veligura}, \citenamefont {Van~Elferen}, \citenamefont {Tombros},
  \citenamefont {Maan}, \citenamefont {Zeitler},\ and\ \citenamefont
  {Van~Wees}}]{veligura2012transport}%
  \BibitemOpen
  \bibfield  {author} {\bibinfo {author} {\bibfnamefont {A}~\bibnamefont
  {Veligura}}, \bibinfo {author} {\bibfnamefont {HJ}~\bibnamefont
  {Van~Elferen}}, \bibinfo {author} {\bibfnamefont {N}~\bibnamefont {Tombros}},
  \bibinfo {author} {\bibfnamefont {JC}~\bibnamefont {Maan}}, \bibinfo {author}
  {\bibfnamefont {U}~\bibnamefont {Zeitler}}, \ and\ \bibinfo {author}
  {\bibfnamefont {BJ}~\bibnamefont {Van~Wees}},\ }\bibfield  {title} {\enquote
  {\bibinfo {title} {{Transport gap in suspended bilayer graphene at zero
  magnetic field}},}\ }\href@noop {} {\bibfield  {journal} {\bibinfo  {journal}
  {Physical Review B}\ }\textbf {\bibinfo {volume} {85}},\ \bibinfo {pages}
  {155412} (\bibinfo {year} {2012})}\BibitemShut {NoStop}%
\bibitem [{\citenamefont {Kolmogorov}\ and\ \citenamefont
  {Crespi}(2005)}]{kolmogorov2005registry}%
  \BibitemOpen
  \bibfield  {author} {\bibinfo {author} {\bibfnamefont {Aleksey~N}\
  \bibnamefont {Kolmogorov}}\ and\ \bibinfo {author} {\bibfnamefont
  {Vincent~H}\ \bibnamefont {Crespi}},\ }\bibfield  {title} {\enquote {\bibinfo
  {title} {{Registry-dependent interlayer potential for graphitic systems}},}\
  }\href@noop {} {\bibfield  {journal} {\bibinfo  {journal} {Physical Review
  B}\ }\textbf {\bibinfo {volume} {71}},\ \bibinfo {pages} {235415} (\bibinfo
  {year} {2005})}\BibitemShut {NoStop}%
\bibitem [{\citenamefont {Lebedeva}\ \emph {et~al.}(2011)\citenamefont
  {Lebedeva}, \citenamefont {Knizhnik}, \citenamefont {Popov}, \citenamefont
  {Lozovik},\ and\ \citenamefont {Potapkin}}]{lebedeva2011interlayer}%
  \BibitemOpen
  \bibfield  {author} {\bibinfo {author} {\bibfnamefont {Irina~V}\ \bibnamefont
  {Lebedeva}}, \bibinfo {author} {\bibfnamefont {Andrey~A}\ \bibnamefont
  {Knizhnik}}, \bibinfo {author} {\bibfnamefont {Andrey~M}\ \bibnamefont
  {Popov}}, \bibinfo {author} {\bibfnamefont {Yurii~E}\ \bibnamefont
  {Lozovik}}, \ and\ \bibinfo {author} {\bibfnamefont {Boris~V}\ \bibnamefont
  {Potapkin}},\ }\bibfield  {title} {\enquote {\bibinfo {title} {Interlayer
  interaction and relative vibrations of bilayer graphene},}\ }\href@noop {}
  {\bibfield  {journal} {\bibinfo  {journal} {Physical Chemistry Chemical
  Physics}\ }\textbf {\bibinfo {volume} {13}},\ \bibinfo {pages} {5687--5695}
  (\bibinfo {year} {2011})}\BibitemShut {NoStop}%
\bibitem [{\citenamefont {Del~Grande}\ \emph {et~al.}(2019)\citenamefont
  {Del~Grande}, \citenamefont {Menezes},\ and\ \citenamefont
  {Capaz}}]{del2019layer}%
  \BibitemOpen
  \bibfield  {author} {\bibinfo {author} {\bibfnamefont {Rafael~R}\
  \bibnamefont {Del~Grande}}, \bibinfo {author} {\bibfnamefont {Marcos~G}\
  \bibnamefont {Menezes}}, \ and\ \bibinfo {author} {\bibfnamefont {Rodrigo~B}\
  \bibnamefont {Capaz}},\ }\bibfield  {title} {\enquote {\bibinfo {title}
  {Layer breathing and shear modes in multilayer graphene: a dft-vdw study},}\
  }\href@noop {} {\bibfield  {journal} {\bibinfo  {journal} {Journal of
  Physics: Condensed Matter}\ }\textbf {\bibinfo {volume} {31}},\ \bibinfo
  {pages} {295301} (\bibinfo {year} {2019})}\BibitemShut {NoStop}%
\bibitem [{\citenamefont {Mayorov}\ \emph {et~al.}(2011)\citenamefont
  {Mayorov}, \citenamefont {Elias}, \citenamefont {Mucha-Kruczynski},
  \citenamefont {Gorbachev}, \citenamefont {Tudorovskiy}, \citenamefont
  {Zhukov}, \citenamefont {Morozov}, \citenamefont {Katsnelson}, \citenamefont
  {Fal’ko}, \citenamefont {Geim} \emph {et~al.}}]{mayorov2011interaction}%
  \BibitemOpen
  \bibfield  {author} {\bibinfo {author} {\bibfnamefont {AS}~\bibnamefont
  {Mayorov}}, \bibinfo {author} {\bibfnamefont {DC}~\bibnamefont {Elias}},
  \bibinfo {author} {\bibfnamefont {Marcin}\ \bibnamefont {Mucha-Kruczynski}},
  \bibinfo {author} {\bibfnamefont {RV}~\bibnamefont {Gorbachev}}, \bibinfo
  {author} {\bibfnamefont {T}~\bibnamefont {Tudorovskiy}}, \bibinfo {author}
  {\bibfnamefont {A}~\bibnamefont {Zhukov}}, \bibinfo {author} {\bibfnamefont
  {SV}~\bibnamefont {Morozov}}, \bibinfo {author} {\bibfnamefont
  {MI}~\bibnamefont {Katsnelson}}, \bibinfo {author} {\bibfnamefont
  {VI}~\bibnamefont {Fal’ko}}, \bibinfo {author} {\bibfnamefont
  {AK}~\bibnamefont {Geim}},  \emph {et~al.},\ }\bibfield  {title} {\enquote
  {\bibinfo {title} {Interaction-driven spectrum reconstruction in bilayer
  graphene},}\ }\href@noop {} {\bibfield  {journal} {\bibinfo  {journal}
  {Science}\ }\textbf {\bibinfo {volume} {333}},\ \bibinfo {pages} {860--863}
  (\bibinfo {year} {2011})}\BibitemShut {NoStop}%
\bibitem [{\citenamefont {Nam}\ \emph {et~al.}(2018)\citenamefont {Nam},
  \citenamefont {Ki}, \citenamefont {Soler-Delgado},\ and\ \citenamefont
  {Morpurgo}}]{nam2018family}%
  \BibitemOpen
  \bibfield  {author} {\bibinfo {author} {\bibfnamefont {Youngwoo}\
  \bibnamefont {Nam}}, \bibinfo {author} {\bibfnamefont {Dong-Keun}\
  \bibnamefont {Ki}}, \bibinfo {author} {\bibfnamefont {David}\ \bibnamefont
  {Soler-Delgado}}, \ and\ \bibinfo {author} {\bibfnamefont {Alberto~F}\
  \bibnamefont {Morpurgo}},\ }\bibfield  {title} {\enquote {\bibinfo {title} {A
  family of finite-temperature electronic phase transitions in graphene
  multilayers},}\ }\href@noop {} {\bibfield  {journal} {\bibinfo  {journal}
  {Science}\ }\textbf {\bibinfo {volume} {362}},\ \bibinfo {pages} {324--328}
  (\bibinfo {year} {2018})}\BibitemShut {NoStop}%
\bibitem [{\citenamefont {Liu}\ \emph {et~al.}(2017)\citenamefont {Liu},
  \citenamefont {Lew},\ and\ \citenamefont {Liu}}]{liu2017observation}%
  \BibitemOpen
  \bibfield  {author} {\bibinfo {author} {\bibfnamefont {Yanping}\ \bibnamefont
  {Liu}}, \bibinfo {author} {\bibfnamefont {Wen~Siang}\ \bibnamefont {Lew}}, \
  and\ \bibinfo {author} {\bibfnamefont {Zongwen}\ \bibnamefont {Liu}},\
  }\bibfield  {title} {\enquote {\bibinfo {title} {Observation of anomalous
  resistance behavior in bilayer graphene},}\ }\href@noop {} {\bibfield
  {journal} {\bibinfo  {journal} {Nanoscale Research Letters}\ }\textbf
  {\bibinfo {volume} {12}},\ \bibinfo {pages} {1--8} (\bibinfo {year}
  {2017})}\BibitemShut {NoStop}%
\bibitem [{\citenamefont {Butz}\ \emph {et~al.}(2014)\citenamefont {Butz},
  \citenamefont {Dolle}, \citenamefont {Niekiel}, \citenamefont {Weber},
  \citenamefont {Waldmann}, \citenamefont {Weber}, \citenamefont {Meyer},\ and\
  \citenamefont {Spiecker}}]{butz2014dislocations}%
  \BibitemOpen
  \bibfield  {author} {\bibinfo {author} {\bibfnamefont {Benjamin}\
  \bibnamefont {Butz}}, \bibinfo {author} {\bibfnamefont {Christian}\
  \bibnamefont {Dolle}}, \bibinfo {author} {\bibfnamefont {Florian}\
  \bibnamefont {Niekiel}}, \bibinfo {author} {\bibfnamefont {Konstantin}\
  \bibnamefont {Weber}}, \bibinfo {author} {\bibfnamefont {Daniel}\
  \bibnamefont {Waldmann}}, \bibinfo {author} {\bibfnamefont {Heiko~B}\
  \bibnamefont {Weber}}, \bibinfo {author} {\bibfnamefont {Bernd}\ \bibnamefont
  {Meyer}}, \ and\ \bibinfo {author} {\bibfnamefont {Erdmann}\ \bibnamefont
  {Spiecker}},\ }\bibfield  {title} {\enquote {\bibinfo {title} {Dislocations
  in bilayer graphene},}\ }\href@noop {} {\bibfield  {journal} {\bibinfo
  {journal} {Nature}\ }\textbf {\bibinfo {volume} {505}},\ \bibinfo {pages}
  {533--537} (\bibinfo {year} {2014})}\BibitemShut {NoStop}%
\bibitem [{\citenamefont {Pereira}\ \emph {et~al.}(2009)\citenamefont
  {Pereira}, \citenamefont {Ribeiro}, \citenamefont {Peres},\ and\
  \citenamefont {Castro~Neto}}]{PhysRevB.79.045421}%
  \BibitemOpen
  \bibfield  {author} {\bibinfo {author} {\bibfnamefont {Vitor~M.}\
  \bibnamefont {Pereira}}, \bibinfo {author} {\bibfnamefont {R.~M.}\
  \bibnamefont {Ribeiro}}, \bibinfo {author} {\bibfnamefont {N.~M.~R.}\
  \bibnamefont {Peres}}, \ and\ \bibinfo {author} {\bibfnamefont {A.~H.}\
  \bibnamefont {Castro~Neto}},\ }\bibfield  {title} {\enquote {\bibinfo {title}
  {Distortion of the perfect lattice structure in bilayer graphene},}\ }\href
  {\doibase 10.1103/PhysRevB.79.045421} {\bibfield  {journal} {\bibinfo
  {journal} {Phys. Rev. B}\ }\textbf {\bibinfo {volume} {79}},\ \bibinfo
  {pages} {045421} (\bibinfo {year} {2009})}\BibitemShut {NoStop}%
\bibitem [{\citenamefont {Dahal}\ \emph {et~al.}(2010)\citenamefont {Dahal},
  \citenamefont {Wehling}, \citenamefont {Bedell}, \citenamefont {Zhu},\ and\
  \citenamefont {Balatsky}}]{dahal2010charge}%
  \BibitemOpen
  \bibfield  {author} {\bibinfo {author} {\bibfnamefont {Hari~P}\ \bibnamefont
  {Dahal}}, \bibinfo {author} {\bibfnamefont {Tim~O}\ \bibnamefont {Wehling}},
  \bibinfo {author} {\bibfnamefont {Kevin~S}\ \bibnamefont {Bedell}}, \bibinfo
  {author} {\bibfnamefont {Jian-Xin}\ \bibnamefont {Zhu}}, \ and\ \bibinfo
  {author} {\bibfnamefont {AV}~\bibnamefont {Balatsky}},\ }\bibfield  {title}
  {\enquote {\bibinfo {title} {Charge inhomogeneity in a single and bilayer
  graphene},}\ }\href@noop {} {\bibfield  {journal} {\bibinfo  {journal}
  {Physica B: Condensed Matter}\ }\textbf {\bibinfo {volume} {405}},\ \bibinfo
  {pages} {2241--2244} (\bibinfo {year} {2010})}\BibitemShut {NoStop}%
\bibitem [{\citenamefont {Nilsson}\ \emph {et~al.}(2006)\citenamefont
  {Nilsson}, \citenamefont {Neto}, \citenamefont {Peres},\ and\ \citenamefont
  {Guinea}}]{nilsson2006electron}%
  \BibitemOpen
  \bibfield  {author} {\bibinfo {author} {\bibfnamefont {Johan}\ \bibnamefont
  {Nilsson}}, \bibinfo {author} {\bibfnamefont {AH~Castro}\ \bibnamefont
  {Neto}}, \bibinfo {author} {\bibfnamefont {NMR}\ \bibnamefont {Peres}}, \
  and\ \bibinfo {author} {\bibfnamefont {F}~\bibnamefont {Guinea}},\ }\bibfield
   {title} {\enquote {\bibinfo {title} {Electron-electron interactions and the
  phase diagram of a graphene bilayer},}\ }\href@noop {} {\bibfield  {journal}
  {\bibinfo  {journal} {Physical Review B}\ }\textbf {\bibinfo {volume} {73}},\
  \bibinfo {pages} {214418} (\bibinfo {year} {2006})}\BibitemShut {NoStop}%
\bibitem [{\citenamefont {Flachi}(2013)}]{flachi2013strongly}%
  \BibitemOpen
  \bibfield  {author} {\bibinfo {author} {\bibfnamefont {Antonino}\
  \bibnamefont {Flachi}},\ }\bibfield  {title} {\enquote {\bibinfo {title}
  {{Strongly interacting fermions and phases of the Casimir effect}},}\
  }\href@noop {} {\bibfield  {journal} {\bibinfo  {journal} {Physical review
  letters}\ }\textbf {\bibinfo {volume} {110}},\ \bibinfo {pages} {060401}
  (\bibinfo {year} {2013})}\BibitemShut {NoStop}%
\end{thebibliography}%

\end{document}